\documentclass[manuscript, screen]{acmart}

\usepackage[american]{babel}
\usepackage[utf8]{inputenc}
\usepackage{todonotes}
\usepackage{multirow,multicol}
\newcounter{nc}

\newcommand{\figstr}{Figure}

\newcommand{\atributosDS}{FeaturesDS}
\newcommand{\IS}{Sustainability Index}
\newcommand{\indicadoresDS}{IndicatorsDS}
\newcommand{\capstr}{Chapter}
\newcommand{\secstr}{Section}

\newcommand{\tabstr}{Table}

\newcommand{\isadigital}{Agro 4.0}
\usepackage{csquotes}

\def\BibTeX{{\rm B\kern-.05em{\sc i\kern-.025em b}\kern-.08emT\kern-.1667em\lower.7ex\hbox{E}\kern-.125emX}}
\begin{document}
%
\title{\isadigital~: A Green Information System for Sustainable Agroecosystem Management}
\acmJournal{TDSCI}


%
\author{Eugênio Pacceli Reis da Fonseca}
\email{eugenio.pacceli@dcc.ufmg.br}

\author{Evandro Caldeira}
\email{evandro@dcc.ufmg.br}

\author{Heitor Soares Ramos Filho}
\email{heitor.ramos@gmail.com}

\author{Leonardo Barbosa e Oliveira}
\authornote{Also a Visiting Associate Professor at Stanford University.}
\email{leob@dcc.ufmg.br}

\author{Adriano César Machado Pereira}
\email{adrianoc@dcc.ufmg.br}

\affiliation{%
  \institution{Universidade Federal de Minas Gerais}
  \streetaddress{Av. Antônio Carlos, 6627 - Pampulha}
  \city{Belo Horizonte}
  \state{Minas Gerais}
  \country{Brazil}
  \postcode{31270-901}
}

\author{Pierre Santos Vilela}
\email{psvilela@faemg.org.br}
\affiliation{%
  \institution{INAES, Federação da Agricultura e Pecuária do Estado de Minas Gerais}
  \streetaddress{Avenida do Contorno, 1.771}
  \city{Belo Horizonte}
  \state{Minas Gerais}
  \country{Brazil}
  \postcode{30110-005}
}

\renewcommand{\shortauthors}{Eugênio Fonseca, Evandro Caldeira, Heitor Ramos, Pierre Vilela, Leonardo B.\ Oliveira, Adriano Pereira, et al.}

\begin{abstract}
Agriculture is one of the most critical activities developed today by humankind and is in constant technical evolution to supply food and other essential products to everlasting and increasing demand. New machines, seeds, and fertilizers were developed to increase the productivity of cultivated areas. The idea of sustainable development became widely known after the United Nations Conference on Environment and Development in Rio de Janeiro, Brazil, in 1992 (ECO-92). It is estimated that by 2050 we will have a population of 9 billion people and the production of food to meet this demand must occur sustainably. To achieve this goal, it is paramount the adoption of sustainable management techniques for agroecosystems. However, this is a complex task due to a large number of variables involved. One of the solutions for the handling and treatment of such diverse data is the use of Green IS. In this work, we adopt a methodology called Indicators of Sustainability in Agroecosystems (\textit{Indicadores de Sustentabilidade em Agroecossistemas} -- ISA), implement an information system based on it and apply Data Science techniques over the gathered data - from 100 real rural properties - to compute which are the most relevant ISA Indicators for the final ISA Sustainability Index Score. As a result, we have developed a set of tools for data collection, processing, visualization, and analysis of the sustainability of a rural property or region, following the ISA methodology. We also have that with only 7 of the 21 Indicators present in ISA we can identify the level of sustainability in more than 90\% of cases, allowing for a new discussion about shrinking the amount of data needed for the computation of ISA, or remodelling the final computation of the Sustainability Index so other Indicators can be more expressive. Users of the solutions developed in this work can identify best practices for sustainability in participating agroecosystems.
\end{abstract}


%

%

%

\settopmatter{printacmref=false}

\maketitle

\pagebreak

\section{Introduction}

Agriculture is one of the most critical activities developed today by humankind. Today the world has about 7 billion people, and agriculture is continuously improving to produce food for all these people. New machines, seeds, and fertilizers were developed to increase the productivity of cultivated areas. The increase in food production reduced hunger, improved nutrition, and prevented new areas from being converted to agricultural use ~\citep{Tilman_2002}. 

Sustainability is when one can work on the present development without compromising the development of future generations~\citep{keeble1988brundtland}. The idea of sustainable development became widely known after the United Nations Conference on Environment and Development in Rio de Janeiro in 1992 (ECO-92). Since then, there have been variations in the definition of sustainability \citep{bebbington2001account, vanMarrewijk2003, glavivc2007review, dempsey2011social}, but they all converge to a definition where sustainability implies a medium- and long-run profitability, as well as agricultural practices with sustainable environmental impacts. Public awareness of the negative impacts of human activity on our environment is at an all times high, and, according to specialists' predictions, no more time can be wasted \citep{climateemergency}

It is estimated that by 2050 we will have a population of 9 billion people and the production of food to meet this demand must occur sustainably. However, achieving this objective is a complex task due to a large number of variables. In addition to a solid model for measuring sustainability, relevant data needs to be well structured for retrieval, storage and analysis. There are also other problems related to understanding the factors that affect sustainability. Therefore, there is a dire need for a system able to: a) characterize, visualize, and analyze collected data; and b) implement smart strategies that can measure sustainability in agroecosystems.

As a solution to this problem, we have designed, developed, and evaluated a sustainability management system in agroecosystems based on data science. The system, dubbed \isadigital~,  provides tools for the collection, storage, analysis, and visualization of sustainability-related information of rural properties. \isadigital~ enables, for instance, different properties to be compared regarding their Sustainability Index. Besides, the system can pinpoint agro-ecosystems with critical levels of sustainability and then suggest managers measures to reverse the situation. It is worth saying that \isadigital~is so-called because it is based on the Agroecosystems Sustainability Index (ISA)~\cite{marzall2000indicadores} methodology. 

\isadigital~ is inserted in a multidisciplinary and multifacet context. First of all, it is an \textit{Green Information System} that supports \textit{Decision Making} with data, information extraction and visualization. The system facilitates the work of groups and entities that seek to improve the sustainability of properties of a given profile. This profile can be, for instance, for the type of product produced. The system makes use of \textit{Data Science}, which enables managers of projects to perform more complex analyzes over a set of rural properties. \textit{Agroecosystems} and \textit{Sustainability} are at the core of the ISA Methodology, that \isadigital~ implements.

This work has two main contributions: (i) the identification of which indicators of the ISA Model are more relevant or expressive for the Sustainability of a rural property, opening a discussion about the amount of data that ISA requires and the possibility to reduce the input (as it is right now, there are hundreds of fields a technician has to fill in order to obtain the final scores); and (ii) a data science-based information system for sustainability management of agroecosystems that allows to:

\begin{enumerate}
    \item collect, to structure and validate data about the sustainability of agroecosystems using the ISA Methodology;
    \item manage information about sustainability and support decision in agroecosystems;
    \item identify and characterize the most relevant factors for sustainability;
    \item perform visualization and analysis over aggregate data in a user-friendly way.
\end{enumerate}

Besides, we have validated \isadigital~ by using data from one hundred rural properties in the state of Minas Gerais, Brazil. Minas Gerais was Brazil's greatest producer of both coffee and milk in 2018. Brazil was the world's greatest producer and exporter of coffee in 2018 and in previous years as well. Brazil was also the worlds' greatest exporter of both beef and chicken in the same year. There were more cattle heads than people in Brazil in the year 2018, according to FAS/USDA.

The rest of this work is organized as follows. In Section~\ref{cha:fundamental} we present the theoretical basis of our work describing concepts of agroecosystems sustainability, and data science. In Section \ref{cha:related} we discuss the related work, concepts of agroecosystems, sustainability and data science. In Section~\ref{ch:aplicacao} we present our platform and its architecture. In Section~\ref{ch:case} we describe a case study based on rural properties located in the countryside of Brazil. Finally, in Section \ref{cha:conclusion}, we conclude our work.


\section{Fundamentals}

\label{cha:fundamental}

In this section, we describe the theoretical fundamentals needed for the comprehension of this paper. Section~\ref{sec:sustainability} describes the main concepts related to agroecosystems sustainability, focusing on the ISA model and methodology. 

\subsection{Sustainability}
\label{sec:sustainability}

One of the current challenges faced by productive systems is balancing sustainable production with societies' needs, or market demands. Certifications are used, on the industrial sectors, to reduce the environmental impacts of such activities and guide the processes involved towards improvements, so they become more efficient lessening their impact on the environment. Some of those certificates are the ISO 14001 and the EMAS\footnote{\url{http://ec.europa.eu/environment/emas/tools/faq_en.htm}}(\textit{Eco-Management and Audit Scheme}), currently being practiced in the European Union. The EMAS is more rigid, precise and yet more reaching\footnote{\url{http://www.emas.de/fileadmin/user_upload/04_ueberemas/PDF-Dateien/Unterschiede_iso_en.pdf}} than the ISO, that being the reason it was picked for implementation by the European Union.

The elaboration of Sustainability Indicators regarding agriculture is a complex task that begins with the definition of parameters to be monitored (soil erosion, soil acidity, production efficiency, among others). The definition of these parameters and the meaning of the indicators can also be influenced by regionality or geography, noting that some parameters cannot be applied uniformly for every region, like, for instance, water salinity~\citep{freebairn2003reflections}. Acknowledging the complex nature of such task (elaboration of environmental indicators) and taking that into account, \citet{Rogmans2016603} and \citet{Coteur201616} proposed guidelines for the definition of such indicators. \citet{Singh2009189} studied 41 methodologies for computing and estimating sustainability indicators, each with recommended scenarios and use cases, citing three: indicators for urban development, environmental vulnerability for cities and indicators for green policies effectiveness.





\subsection{Green Information System}

The Green Information System - or Green IS - label has a broad definition. According to Watson et al. [2008], it refers to an Information System that supports or enables sustainable initiatives and addresses environmental issues, aiming to reduce an activity's impacts on the environment. The Green IS thus has an indirect impact on the environment, through the positive impact it has on the activity it supports \citep{greenisconcepts} \citep{greenisantecedents}.

In literature, a related term is Green Information Technology, or Green IT, which is an Information Technology practice or study focused on reducing the first order impacts of IT activities on the environment. Examples of Green IT practises are introducing energy-efficient hardware to an IT operation or providing a sustainable framework to handle the disposal of IT equipment. Green IT is often related to hardware; it is related to software when the software focuses on mitigating the immediate impacts of an IT activity \citep{greenisconcepts} \citep{senioris} \citep{greenisagenda}.

Examples of Green IS, referring to Information Systems that indirectly affect the environment by improving the sustainability of activities those give support to, can be: IS that aims to provide support to supply chains, optimizing routes and transportation; IS that monitor environmental variables such as water and energy consumption, waste, emissions, toxicity and carbon footprints of an industry, among others \citep{watson2008}.

We argue that the system presented in this work is a Green Information System, precisely because it is used to register and monitor, yearly, variables such as water contamination and quality, usage of agrotoxics, soil quality and contamination, size and status of legal preservation areas, waste management, among others, of an agricultural economic activity, so technicians can help producers in improving their business' sustainability, lowering their impacts on the environment.

\subsection{Indicators of Sustainability in Agroecosystems}
\label{subsec:isa}

A way to evaluate the sustainability of rural properties and farming businesses is the System of Sustainability Evaluation (SAS)~\citep{rodrigues2013sustentabilidade}, applied to measure the sustainability of ethanol and sugar cane productive businesses on the state of São Paulo. This methodology, despite being well detailed and accurate in some aspects, regarding air quality measurements, for instance, is not generic enough to apply to rural businesses with other productive profiles.

The ISA project is an initiative of the State Secretary of Agriculture, Pecuary and Supplies of Minas Gerais (SEAPA), Brazil. The methodology proposed by the ISA project allows a detailed check on a target rural property, highlighting a compounded analysis of their production systems, information management, water and soil qualities, natural habitat preservation, employment conditions and quality for the workers, among other characteristics. The ISA Platform is accessible on \url{http://www.epamig.br/projeto-isa/}. Environmental Sustainability is also a factor of economic interest for the municipalities that house rural businesses. For instance, some Brazilian states, such as Paraná, São Paulo and Minas Gerais reward municipalities that take good care of their natural environment by providing tax benefits, as measured by the system~\citep{hempel2009importancia}.


ISA is composed of 21 indicators which values are in the interval $[0; 1]$~\citep{marzall2000indicadores, ferreira2012indicadores}. To help with better analysis and information extraction, those indicators are grouped by sub-indexes, detailed in table \ref{tab:subindexesindicators}.

\begin{table}
  \caption{Sub-Indexes and Indicators of ISA}
  \label{tab:subindexesindicators}
  \begin{tabular}{l l}
    \toprule
    Sub-Index & Indicators computed for the score \\
    \midrule
    \multirow{4}{*}{1. Economic Balance} & Productivity \\
        & 1. Income Diversification \\
        & 2. Assets Development \\
        & 3. Degree of Indebtedness \\ \hline
    \multirow{3}{*}{2. Social Balance} & 5. Basic Services Availability \\
        & 6. Scholarship \\
        & 7. Work/Employment Quality \\ \hline
    \multirow{4}{*}{3. Business Management} & 8. Business Management \\
        & 9. Information Management \\
        & 10. Residues Management \\
        & 11. Work Security indicators \\ \hline
    4. Soil Productive Capacity & 12. Soil Fertility indicator \\ \hline
    \multirow{2}{*}{5. Water Quality} & 13. Water Quality \\
        & 14. Contamination Risks (concerning the usage of pesticides) \\ \hline
    \multirow{3}{*}{6. Handling of the Production Systems} & 15. Soil Degradation Evaluation \\
        & 16. Conservation Practices Adoption \\
        & 17. Roads Quality indicators \\ \hline
    \multirow{4}{*}{7. Ecology of the Rural Landscape} & 18. Native Vegetation \\
        & 19. Permanent Preservation Areas \\
        & 20. Legal Reserve Area \\
        & 21. Landscape Diversification indicators \\
    \bottomrule
  \end{tabular}
\end{table}

\begin{figure}[!htb]
    \centering
    \includegraphics[width=0.7\textwidth]{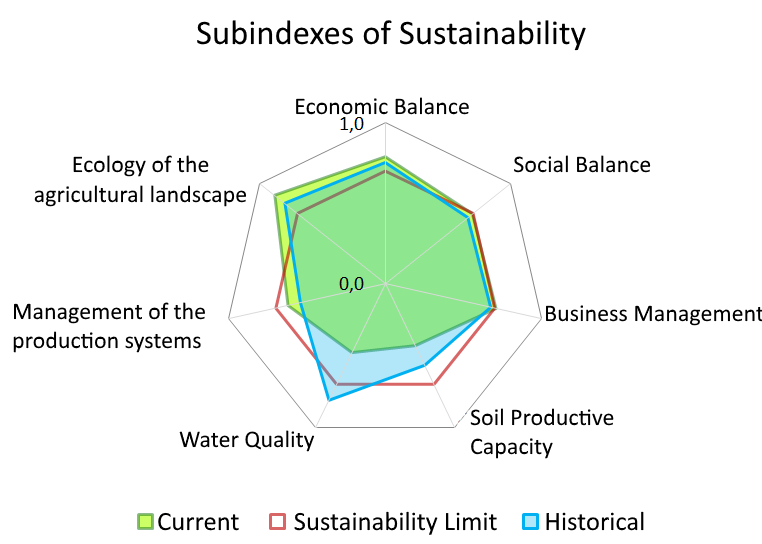}
    \caption{ISA - Sustainability sub-indexes in a sample ISA  questionnaire implemented in Microsoft Excel}
    \label{fig:isaSubindice}
\end{figure}

\begin{figure}[!htb]
    \centering
    \includegraphics[width=0.7\textwidth]{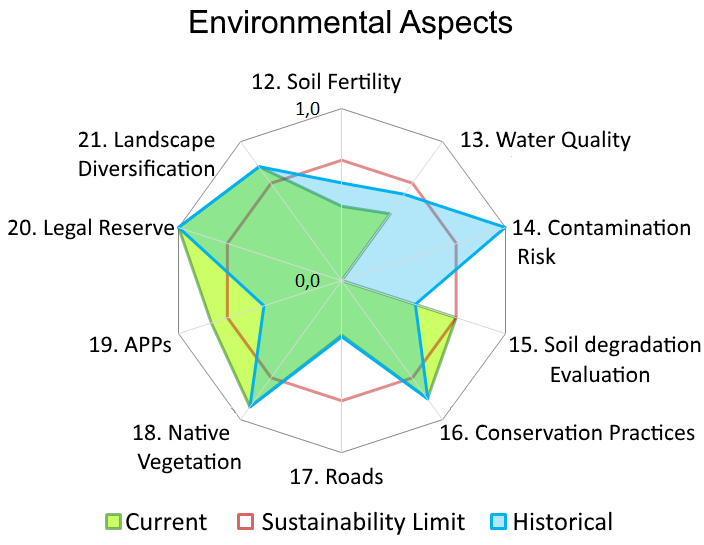}
    \caption{ISA - Environmental Aspects, generated by a sample ISA  questionnaire implemented in Microsoft Excel}
    \label{fig:isaAmbiental}
\end{figure}

The ISA Platform allows for the synthesizing of all that data into  informative charts, one kind exemplified in Figure~\ref{fig:isaSubindice}.
Details regarding the Environmental Aspects set of indicators can be seen in Figure~\ref{fig:isaAmbiental}. In this sample, some  indicators are plotted together to give an user an overview of  environmental characteristics of a rural property, such as its Water Quality, Soil Degradation, Permanent Preservation Areas, Legal Reserve Areas, among others.

Figures~\ref{fig:isaSubindice} and~\ref{fig:isaAmbiental} show the \textit{Sustainability Limit} minimum acceptable value (0.7, represented in those radar charts by red heptagons). With this limit, it is easy to identify which aspects need to be improved to achieve better sustainability. The historical value, in blue, allows for comparisons along time, so advances or setbacks are identifiable.

\begin{figure}[!htb]
    \centering
    \includegraphics[width=0.7\textwidth]{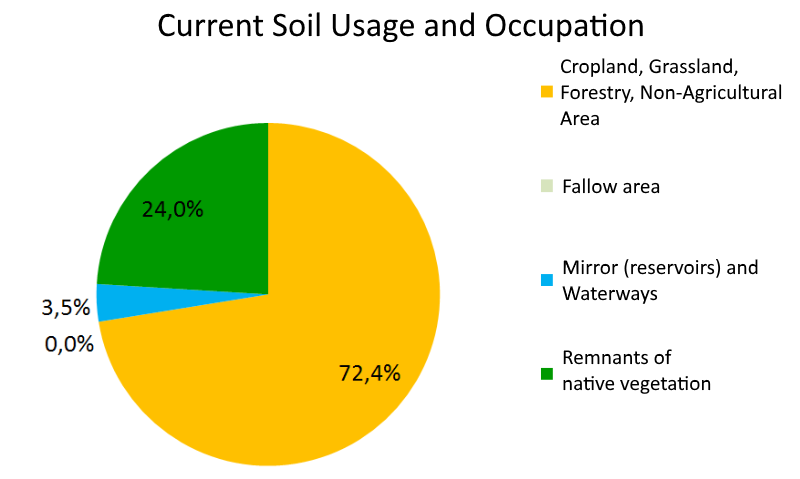}
    \caption{ISA - Soil use occupation generated by a sample ISA  questionnaire implemented in Microsoft Excel}
    \label{fig:isaSolo}
\end{figure}

Figure~\ref{fig:isaSolo} depicts details of the Soil Usage and Occupation of the sample rural property. The figure shows how soil usage is distributed into different crops for a rural property, along with the percentage of their layers of water. ISA collects data other than necessary to compute the Indicators and Sub-Indexes ~\citep{marzall2000indicadores, ferreira2012indicadores}.

\section{Related Works}
\label{cha:related}


The article \citep{bdagriculture} made a review of the literature for big data applications in farming and agriculture. Thirty-four articles were analysed for the tools they used and the problems they tackled. The authors note that s although Big Data is quite successful and popular as a domain, there are still very few cases of its appliances on agriculture, especially on small farming, as the numbers of scientific publications and commercial initiatives show. The authors note that the five Vs of Big Data - Volume, Velocity, Variety, Veracity and Valorization - \citet{chi}~et al. are often misunderstood as people value Volume over the rest, which is equally important. The authors also point out that the sources of data for the solutions studied are very varied: drone images, governmental institutions, weather sensors, historical information, surveys, the web, among other sensors of different natures. The review documents the most used Big Data tools, including algorithms, databases, GIS systems, statistical tools, among others. Machine learning is often used in predictions, and database solutions are very varied. The survey identifies problems that may be slowing down the pace of Big Data in agriculture, such as: privacy issues raised by farmers - regarding the ownership of the data -, security and accuracy doubts, the possibility of the creation of monopolies as valuable data is collected and concentrated by complex solutions, the access to ground information by the team behind the answers, among others. It is also noted that there is a gap of expertise and access to infrastructure in third world countries and small farms. The authors note that many farmers in all parts of the globe are organizing in cooperatives or communities, a move that empowers them and increases sharing of information and data, possibly opening new windows to introduce Big Data into their operations (in fact, the data analytics tool presented by our work was implemented on ground by FAEMG's affiliated cooperatives). The analysis of data by experts can help guide farmers, such as in one of the surveyed works \citet{giller} et al., 2011, in which the analysis of crops' responses to fertilisers allowed the farmers to manage better which fertilisers to use. The article ends by stating that Big Data has the potential to boost productivity and the development of smarter farming, allowing for an increase in production in an environmentally friendly way.


Article \citep{analyticsagriculture} highlights the growing importance of data analytics and informatisation in agriculture, and how the introduction of Big Data in that sector in the United States is reshaping market relations between companies and farmers. Some of the main challenges agriculture faces nowadays, as pointed by the article, is sustainability. What differentiates the current paradigm of agriculture, called Precision Agriculture, from the traditional Conventional Agriculture, is its an emphasis on data collection and analysis to guide decision making and the overcome of challenges. Soil Fertility, over fertilisation, water contamination, water availability and greenhouse gasses emissions are some of the sustainability challenges pointed by the authors of the work. All those items are taken into consideration by the ISA Methodology. Traditional farm supplier firms are equipping themselves with Data Analytics solutions as part of their commercial arsenal, and those solutions involve the systematic analysis of data to provide valuable information to their clients. The article points out how significant players, as well as startups, are making considerable investments in Data Analytics and Big Data solutions, seeing the potential of growth in this segment.


The article \citep{bussint} implements and argues for Business Intelligence - BI - models that factor in the sustainability of a company. In the work, a Sustainability dimension is added to an existing BI model, considering the economic, social and ecologic sustainability dimensions of an exemplary generic corporate business. An example data model for monitoring Sustainability Projects through BI is also shown and discussed. The author argues that the management of corporate sustainability should rely on BI, as such a tool can provide valuable information for analysis. The author also argues that sustainability - all dimensions: economic, social and ecologic - should be part of a corporate business strategy, and so corporate data should necessarily include sustainability data, and a BI model should reflect that.


The article \citep{organizationaladoption} studies how institutional pressures influence the adoption of Green IS in organisations of different nature. The authors also used a classification of Green IS/Green IT that identifies three groups, based on the contribution the deployment of such technologies has in an organisation. Pollution prevention Green IS the adoption of IS to reduce the pollution caused by other activities of the organisation. Product stewardship Green IS is the adoption of IS to enhance the lifecycle management on the supply chain. Sustainable development Green IS is the adoption of IS to transform business activities, reducing their impact on the environment. A survey involving 75 organizations is done and analysed, resulting in a suggestion that both mimetic - the imitation of behaviour of other organizations, such as partners or competitors, that resulted in success for them - and coercive - regulations, contracts, market demands - pressures are essential and often result in a firm implementing Green IS to mitigate it's activities impacts on the environment. 


The authors of \citep{senioris} begin the article by comparing the impacts of information technology and systems on the environment, and then dividing those into two categories: First order impacts are the negative impacts of using and disposing of information systems on the environment, the effort of mitigating those is called Green IT (Green Information Technology); Second order impacts are the positive impacts of using Green IS (Green Information System) as a tool to improve the sustainability of an operation, activity or business. The article studies the reasons and results behind the adoption of Green Information Systems and uses Melville's Belief-Action-Outcome framework (BOA) to evaluate the impacts of the adoption of Green IS in multiple business environments. The authors surveyed 508 managers from various businesses in Malaysia and concluded that the managers' perception and attitude to Green IS as well as coercive pressure (by regulatory bodies, market or business partners) to the firms pushing them to become more environmentally friendly play a big part in the adoption of Green IS. The article also suggests that the adoption of Green IS had a positive impact on the environmental performance of the firms. 

%

An Information System tackling environmental sustainability issues allows managers involved in the productive chain to make more qualified decisions, resulting in benefits regarding the social, economic and environmental aspects of their activities. Those systems, when properly implemented, can bring advantages to the groups that use them~\citep{malhotra2013spurring}. An example of that is the usage of an Information System in the management of energy, resulting in costs reduction. Another use case would be the deployment of sensors in a project for more efficient irrigation systems that could consume less water and energy~\cite{nikolidakis2015energy}.

A general evaluation of systems employed to help on the measurement of sustainability of agricultural and farming properties happened in Denmark~\citep{de2016assessing}. Solutions for the assessment of sustainability - based on indicators - were compared regarding the process and complexity of employing them. More than 40 solutions were evaluated, and only 4 of them met all the desirable criteria and took into account the environmental, social and economic dimensions of sustainability. The RISE ~\citep{hani2003rise} solution was the one with the best results, and it is used to measure the sustainability of farms. The experiment concluded that the usability, complexity of the solution, language use and meeting the expected use value - by developers and farmers - of the information outputted by the solutions are factors that are weighted for the adoption or rejection of the solution.

The AESIS (\textit{Agro-Environmental Sustainability Information System}) was initially applied to organic agriculture and then expanded to other crops~\citep{pacini2011aesis}. That solution comprises many subsystems that generate environmental indicators for each interest point. They formulated possible answers to sustainability questions, together with critical points for the agricultural sectors of the local economic and agroecological zones, identifying thresholds for indicators and setting systems of management with the proper political parameters. This format is similar to ISA~\citep{ferreira2012indicadores}, but here the indicators are divided into subgroups, the critical threshold is the same for all indicators and the actions for tackling the discovered issues are defined in the adequation plan.

SAFE~\citep{van2007safe} (\textit{Sustainability Assessment of Farming and the Environment} ) structures the information regarding an agro-economic system in a hierarchical manner, to evaluate its sustainability. Three levels, called Portion, Farm and Landscape are defined. That framework also aims to explore the agroecological system's data in a more generalist way to obtain a more concise result of its stipulated sustainability. On the environmental aspect, they take into account data labelled by the groups Air, Soil, Water, Energy, and Biodiversity. On the economic perspective, the financial viability of the business is factored in. For the social issue, food production quality and safety, workers' and families quality of life, social and cultural acceptance of the activity are factored in.

Regarding the adoption of Information Systems in large properties, and in large-scale, three different types of systems are identified in~\citep{kropff2001systems}. The first type is responsible for the prediction of future land uses, based on the extrapolation of current tendencies. The system employs measurements verified in the past to identify future states. This process requires quality and precise measurements~\citep{shaw2016characterising} so that they can create simulations with an acceptable degree of trust. The second type is focused on extensive research to define the types and possible land usages. Initially, the methodology performs studies of the biophysics of the system. The land usage optimization is then made by taking into account all the objectives aimed by the employment of those lands. The third kind of system aims at the identification of policies that benefit certain and specified land usages. The definition of the objectives and specific land usages can be performed by taking into account the financial market to determinate future demands and products on the rise~\citep{de2017investigating}.

The low usage of Information Systems by farmers and other rural properties owners could be explained by the immediate economic impact produced by the adoption of the technologies. Beyond the economic factor, it is also noted that age (of the people whose technological solution is aimed at), educational level and the size of the rural properties are also important factors that weight in the adoption or rejection of new technologies~\citep{mittal2016socio}. In the year 2000, it was predicted that industry restrictions and environmental regulations would force the adoption of support technology by farmers and other rural properties owners~\citep{thysen2000agriculture}. Besides restrictions and resistance by smaller producers ~\citep{mittal2016socio}, the usage of FMIS (Farm Management Information Systems) is indispensable for high precision agriculture~\citep{nikkila2010software} (High precision agriculture defined as in \enquote{eletronic monitoring and control applied to data collection, processing and usage for support in decisions regarding the temporal and spacial allocations of supplies for crops}~\citep{bongiovanni2004precision}).

\citet{fountas2015farm} analyzes 141 international FMIS packages, grouped into 11 categories. Commercial solutions from France (10 solutions), Germany (16 solutions), Italy (16 solutions), The United States (62 solutions) and Canada (4 solutions) were evaluated. The table \ref{tab:fmispackagetable} shows all the functions used to map and group the 141 researched packages.

\begin{table}
  \caption{Categories for FMIS packages by \citet{fountas2015farm}}
  \label{tab:fmispackagetable}
  \begin{tabular}{l p{11cm}}
    \toprule
    Category & Description \\
    \midrule
    Field operations & Farm activities registries. \\ \hline
    Better practices & Management of the employment of better practices. \\ \hline
    Finances & Costs estimates for all the activities of the rural property. \\ \hline
    Inventory & Management and monitoring of the inventory used at the crops. \\ \hline
    Traceability & The use of techniques such as labelling to identify the output of each sector of the property as well as to facilitate the rastreability of the allocation of supplies, equipment, and workers. \\ \hline
    Reports & Production, equipment, management and other reports. \\ \hline
    Local specific & Aims to reduce or optimize the quantity of data, improving the generated results. \\ \hline
    Sales & Orders, accounting, supplies, equipment, and Personnel Management. \\ \hline
    Machinery management & Breakdown of equipment usage. \\ \hline
    Human Resources & Management of the workforce. \\ \hline
    Quality Assurance & Monitoring of the production processes and evaluation of those according to legislation. \\
    \bottomrule
  \end{tabular}
\end{table}

Also, during the same work ~\citep{fountas2015farm}, it was stipulated that 75\% of the solutions are developed for personal computers, 10\% only worked on mobile platforms, 9\% are developed as web systems only, and only 6\% of the solutions provide modules for both the mobile and web platforms. The final step of the work was to group the analyzed systems into four categories, described in table \ref{tab:fountastable}.

\begin{table}
  \caption{Categories for the systems identified in \citet{fountas2015farm}}
  \label{tab:fountastable}
  \begin{tabular}{l p{11cm}}
    \toprule
    Category & Description \\
    \midrule
    Basic & Presents intermediate features for reports and sales and basic features for inventory, human resources, and field operations. \\ \hline
    Sales oriented & Presents close to intermediate features regarding reports, finances, sales, inventory, human resources, and field operations. This work presents basic features regarding traceability. \\ \hline
    Local specific & Presents advanced features regarding field operations and local specific activities, features close to intermediate regarding reports and basic features for sales and promoting better practices. \\ \hline
    Complete & Presents advanced features regarding field operations, inventory, finances, and local specific activities; Intermediate features regarding machinery, human resources, reports and traceability; Basic features for sales and promoting better practices. \\
    \bottomrule
  \end{tabular}
\end{table}

The use of sensors facilitates the automation of various processes in agricultural properties.~\citet{abbasi2014review} analyzes different use cases for wireless sensors in signals monitoring. Some of the kinds of signals that can be monitored are temperature, humidity, rain, water levels, conductivity, salinity, hydrogen, $CO_{2}$, winds' speed and direction, atmospheric pressure, among others. Acquisition costs, network types and their use scenarios are also analysed.

Efficient management of water resources can also be achieved with the employment of auxiliar Information Systems. ~\citep{kim2008remote} describes the implementation of a wireless network - composed of wireless sensors and specialized software - for the control of a precision irrigation system. Six sensors stations were installed and distributed on the targetted field following a soil's properties map. Periodic samplings produce data sent to a processing center. That central unit analyses the situation and decides on irrigating specific points (georeferenced by \textit{ sprinklers}) on the field at a given time or not.

Precision agriculture is not solely defined by the adoption of precise tools, as its implementation has other impacts on the way the farms work, and how the farmers labor. Precision agriculture changes the main practices and laboring methods on a rural property. It employs a diverse range of technologies, like GPS (Global Positioning System) and GIS (Geographic Information System), the first is used for the elaboration of the topographies of the rural properties, and precisely positioning sensors. The later can be implemented as a georeferenced database that stores relevant information regarding soil, and its relief, for example. The monitoring of production can be executed using sensors logically scattered along a field, remote sensing can provide satellite images and other area data for the identification of problems on crops, and with that sort of data and others collected by sensors, the exact quantity of nutrients or defensive chemicals can be administered to target areas~\citep{aubert2012enabler}.

\section{The Application}

\label{ch:aplicacao}
The ISA Methodology was conceived by the Enterprise for Agropecuary Research of Minas Gerais (\textit{Empresa de Pesquisa Agropecuária de Minas Gerais} - EPAMIG ) and it's original implementation consisted of using Microsoft Excel for the questionnaire collecting the data (all the fields named in appendix \ref{ch:appendix}) for each property, with a complex Excel worksheet that computed all the indicators, sub-indexes and other immediate results on the go~\citep{marzall2000indicadores, ferreira2012indicadores}. The Excel approach was deemed not suitable for analysis extrapolating more than one property. The sheets could slightly vary in format and patterns of form filling, not all data could be guaranteed to be properly validated, and there was no tool to aggregate and extract information regarding a set of properties, limiting the analysis to technicians manually extracting information from various sheets with hundreds of fields each.

The system developed in this work offers a centralized data collection and visualization approach for the ISA Methodology. Solutions were developed to make it easier for agrarian technicians to collect the ISA questionnaire data when they visit a target rural property and later interpret the results obtained. Data from various properties are collected, structured and processed to generate different reports and diagnostics for each participating rural property.

The \isadigital~system was deployed and tested with participating rural properties from the state of Minas Gerais, Brazil, from the second semester of 2016 until the first semester of 2017. The Federation of Agriculture and Livestock of the State of Minas Gerais (\textit{Federação da Agricultura e Pecuária do Estado de Minas Gerais} - FAEMG) was the institution that applied the system developed in this work in real properties, through their technicians.

The following sections describe the solutions and modules that compose the system.

\subsection{Architecture}\label{sec:app:arquitetura}
 
\begin{figure}
    \centering
    \includegraphics[width=0.7\linewidth]{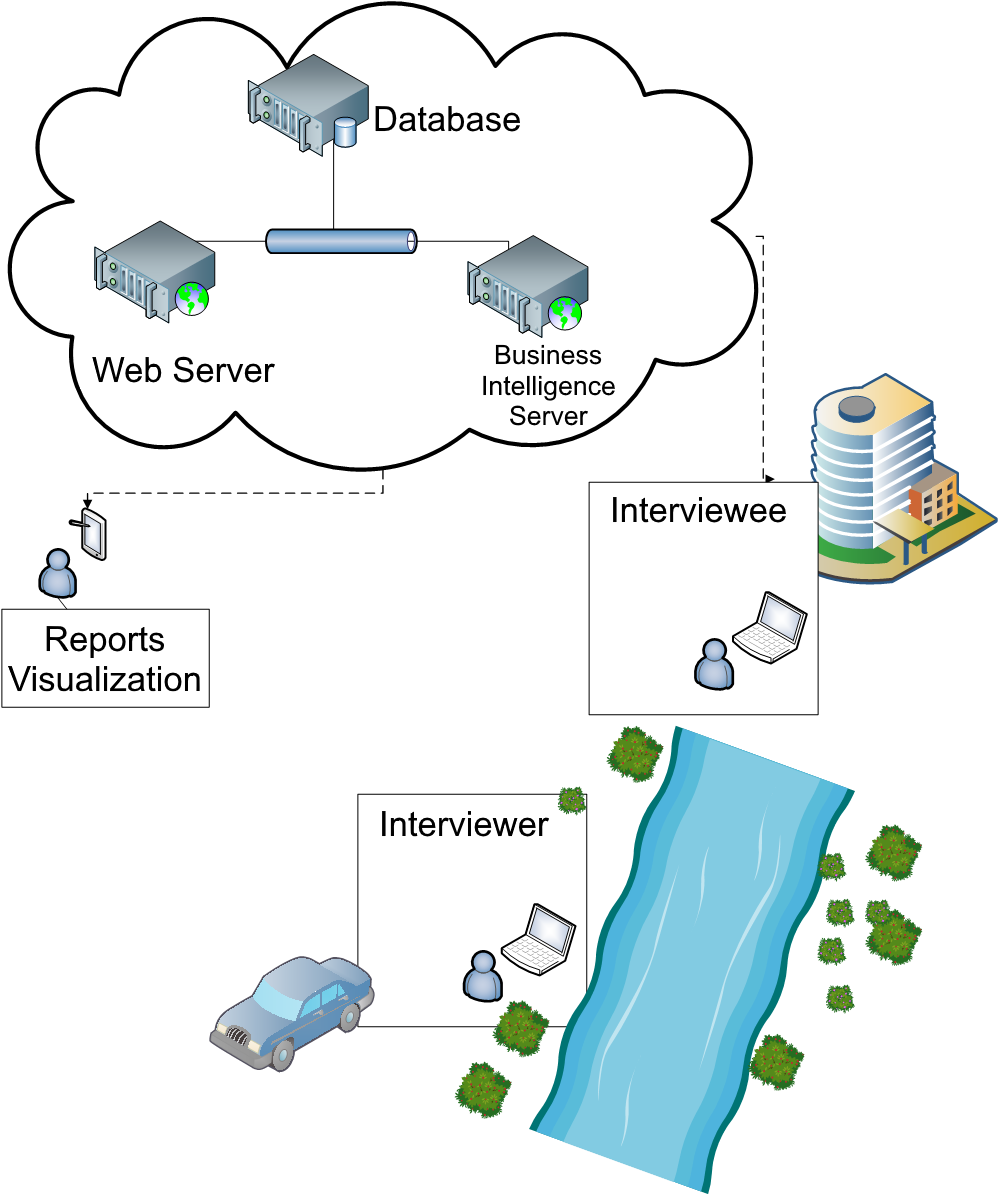}
    \caption{System architecture}
    \label{fig:architeture}
\end{figure}

\begin{figure}
    \centering
    \includegraphics[width=1.0\linewidth]{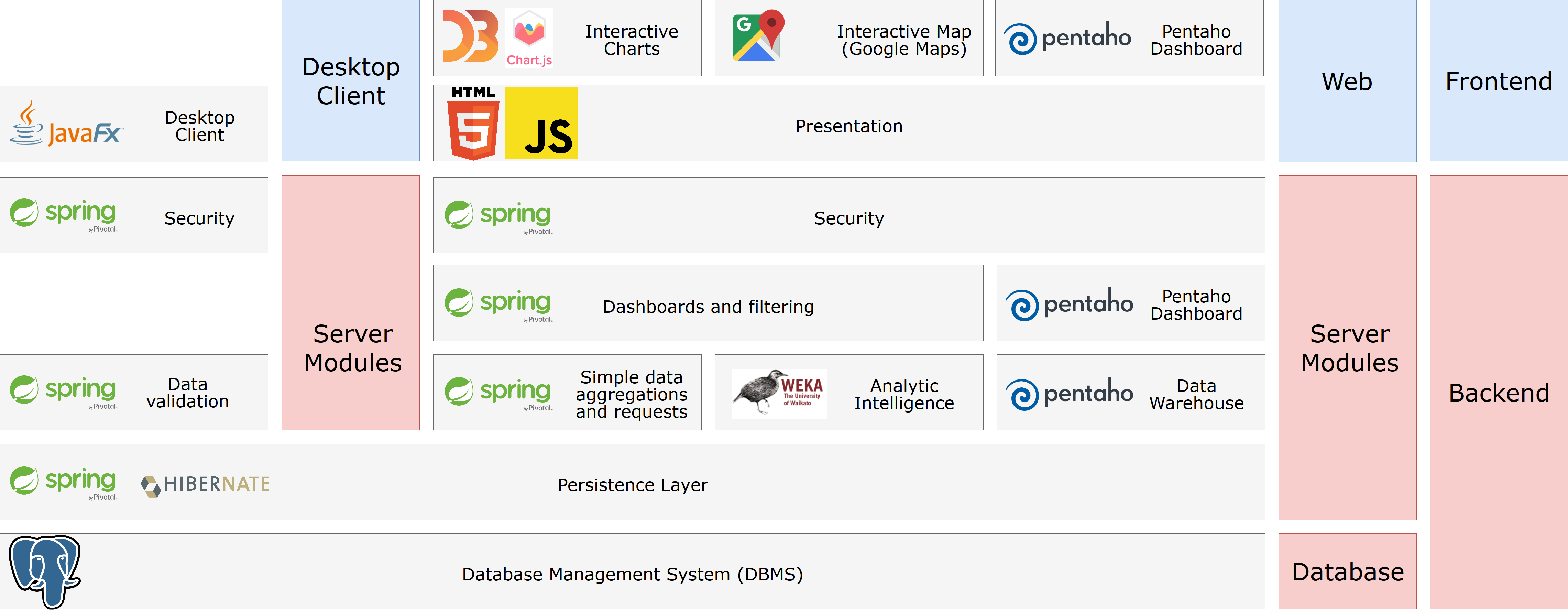}
    \caption{\isadigital~ architecture: software stack}
    \label{fig:architetureStack}
\end{figure}

Figure~\ref{fig:architeture} shows the architectural aspects of the \isadigital~solution. \isadigital~is a multiplatform system, featuring independent components for different tasks (data collection, storage, processing, mining, and visualization). Figure \ref{fig:architetureStack} later shows the software stack of the system, in which each block depends on the ones bellow them in the same column. Appendix \ref{ch:appendix2} features a detailed UML diagram for the most relevant human interaction and information delivery processes of the \isadigital~system, said diagram is also divided into smaller process diagrams in the following modules' subsections present in this section of the article.

The data is primarily collected on the property through an interview made by an able technician with a representative of the rural property. The interview is intended to collect all the data required to fulfil an ISA questionnaire. The questionnaire is composed of hundreds of fields (detailed in appendix \ref{ch:appendix}), many of them requiring technical and precise data (i.e. the pH acidity level of the water streams on the property, the proportion of metals collected by soil sampling, age and training level of employees, among others) ~\citep{marzall2000indicadores, ferreira2012indicadores}, thus the interview process can take more than a day. A ISA questionnaire refers to the state of a rural property in a given year, however, in \isadigital, each questionnaire is also linked to a project.

An \isadigital~project is a set of properties grouped by geographical regions and closed time windows. In the questionnaire, to be filled through a Java desktop client, the Interviewer attributes it to a project, date and the property the questionnaire refers to. A project is also always associated to an institution, that is responsible for it.

The module that receives and validates input data is a Java 8 and JavaFX desktop application. This client sends formatted and validated data to a remote  database, or stores it in the user's (agrarian technician) disk - with the classic save to file button approach - until they decide it is complete and appropriate to be sent. This client also allows the user to download data sent in the past and update it if they wish.

The data is collected in the format of a questionnaire, with multiple tabs and fields. This data collection client is not a web application because some or many rural properties may not be equipped with an Internet connection. The database that receives the data collected by the system is used by the data warehouse, mining and visualization modules for the generation of reports and analysis.

When the questionnaire is ready to be sent, the technician does so, and the data is thus sent a Java EE server (built with Spring MVC and Hibernate, running on Wildfly 15), which stores it in a PostgreSQL database shared with all the other modules. The other modules, running on the same server, are now able to generate the reports, perform the intelligence analysis solutions, feed the data to the data warehouse module and aggregate the questionnaire's header data (date, water basin and municipality) in the sets available for user-guided filters. The server also allows the technicians to fill the Adequation Plan through the browser (the plan to increase a property's sustainability index, as described by the ISA model~\citep{marzall2000indicadores, ferreira2012indicadores}) and the representatives to see it.

One of the advantages of this new solution, when compared to the traditional method of application of the ISA methodology (Excel sheets manually filled by agrarian technicians), is the centralized and structured data storage on the \isadigital~server, using a Relational Database Management System (RDBMS), while also having additional rules for data integrity validated by the persistence layer of the Java EE server (Hibernate). This advantage allows for the automatic processing of data, creation of more complex and trustable analysis and also makes the system more flexible to the additions of new modules in the future.

Reports are generated for each questionnaire that is collected and sent to the server on the go. A user with sufficient permissions can also request dynamically generated reports for a collective of properties: it is possible to filter sets of properties by specific characteristics (location, year of data input, associated institutions, and others) on the fly. Those reports can be accessed through any modern web browser that supports HTML5. The charts and graphs are plotted using either D3.js or Charts.js - depends on the report -.

A Data Warehouse (DW) solution is used to provide more complex and sophisticated data analysis to Managers. The Data Warehouse module is implemented with the data analytics serve Pentledaho CEAddit, by Hitachi Software. ionally, with tr calhis software, more capabilities for further integration with other databases, or future hypothesis and cause-effect investigations, are available.

A software stack overview can be seen in \figstr~\ref{fig:architetureStack}. Each grey block on the image represents a layer that depends on the blocks layered bellow them on the same column. The top blocks, at the same level of the "Frontend" blue block, represent the browser and desktop clients stack. The front-end blocks depends on the back-end stack to exchange data. All access is regulated by a security layer, managed by the Spring Security software, and all data ultimately comes and goes to the central database, managed by Postgres.

Through the web application module, allowed users can visualize useful information for various sets of properties, interactively adjusting and filtering by different properties' or questionnaires' characteristics. They can also request reports for a single questionnaire. All those reports and information aggregation are the results of processing the questionnaires sent by the Interviewers, for each property and year. The system provides three dashboards for data aggregation and visualization. Each dashboard is dedicated to different kinds of information: obtained by simple aggregation and processing; obtained by data mining; generated by the data warehouse solution. A list of each report generated by each questionnaire received is also available so that the user can request for its results, and a map for each questionnaire and property, showing the locations to which those refers to (a form in the questionnaire asks for the properties' coordinates, they are used to plot the property in that map) is also present. The Interviewers can visualize, aggregate and query for all the properties' questionnaires they sent, while the Interviewees or property managers can only see the reports and data related to their properties.

\begin{table}
  \caption{Actors of the \isadigital~ system}
  \label{tab:actorsisadigital}
  \begin{tabular}{l p{11cm}}
    \toprule
    Actor & Description \\
    \midrule
    The Interviewer & Actor who is most of the times a trained agrarian technician, is responsible for filling - through an interview with people responsible for a property - and sending the questionnaires for each property. The technician interviewer is also responsible for formulating a diagnosis and suggesting actions to be taken, by a responsible for the rural property, to further increment the sustainability indicators, after analyzing the results for a questionnaire processed by the system. Registers Interviewees to the system. \\ \hline
    The Interviewee & The person interviewed during the data collection stage, aiding the technician to fill in the forms of the questionnaire for the property they are being questioned about. They are also receiving a diagnosis and suggestions by the technician, by accessing the system later and checking its generated reports, results and technician notes for their property. \\ \hline
    A Project Manager & An Actor that may be, for example, an Agrarian Engineer -, that can visualize the gathered data and its results for a (or multiple) sets of properties in one or more projects they manage. Can add new Interviewers to the system. \\ \hline
    The Institutional Manager & A representative of a cooperative, union or other kinds of associations of rural properties, has the ability to visualize the gathered data and its results for properties and projects linked to their institution(s). Can register new Institutional Managers, Project Managers and Interviewers to the system. \\
    \bottomrule
  \end{tabular}
\end{table}

The \isadigital~system involves four kinds of human actors, described in table \ref{tab:actorsisadigital}. All human actors have login credentials in the system. The Managers and Interviewees mainly use it to visualize data and information, the Interviewers use to send the questionnaires collected through the client, visualize their reports and write Adequation Plans. the Managers are able to visualize and aggregate more data and information than the technicians and interviewees, which are limited to reports and information regarding their questionnaires. The managers can also add new user to the system and set them to act like one of the four actors. The technicians are the only actors writing ISA Methodology related data to the system other than the automatic solutions for data analysis. They are also the only actors that interact with the system outside of the web browser (they use the Java desktop client to send data).

The workflow of the Interviewer comprises of three steps. For the first step, the Interviewer goes to a designated rural property and applies the questionnaire, filling it in an interview with a representative of the said property. Once that task is done and there is internet available, the second step comes in: the Interviewer submits the questionnaire to the \isadigital~server and checks later, through the browser, the report the system generated for the questionnaire. The third step consists of the Interviewer writing the Adequation Plan, a part of the ISA Methodology. The interviewee can access the Adequation Plan written by the technician on the web, it contains orientation on how to increase the sustainability of a property given the results of the information collected through the questionnaire.

A Project has a Project Manager, an actor responsible for the management and supervision of a project. Managers can register, associate or remove the Interviewers from the projects they manage, they can also visualize reports of questionnaires associated to the Projects they manage. They have access to the same dashboards the Institutional Managers have, but limited to display data and information from questionnaires associated to their Projects. A Project can have more than one Project Manager.

The Project and Institutional Managers are responsible for overseeing the application of the ISA Methodology on the targeted rural properties for the Projects they manage. During the application of the ISA Methodology through the \isadigital~system, those professionals can monitor the work of the technicians by checking the reports and Adequation Plans written for each questionnaire submitted. They can also analyze and find patterns on the reports for the properties of a region or project, and perform other kinds of data interpretation they wish through the dashboards they have access to. Managers also manage the users on the system and the allocation of Interviewers to Projects.

\begin{figure}
    \centering
    \includegraphics[width=.85\linewidth]{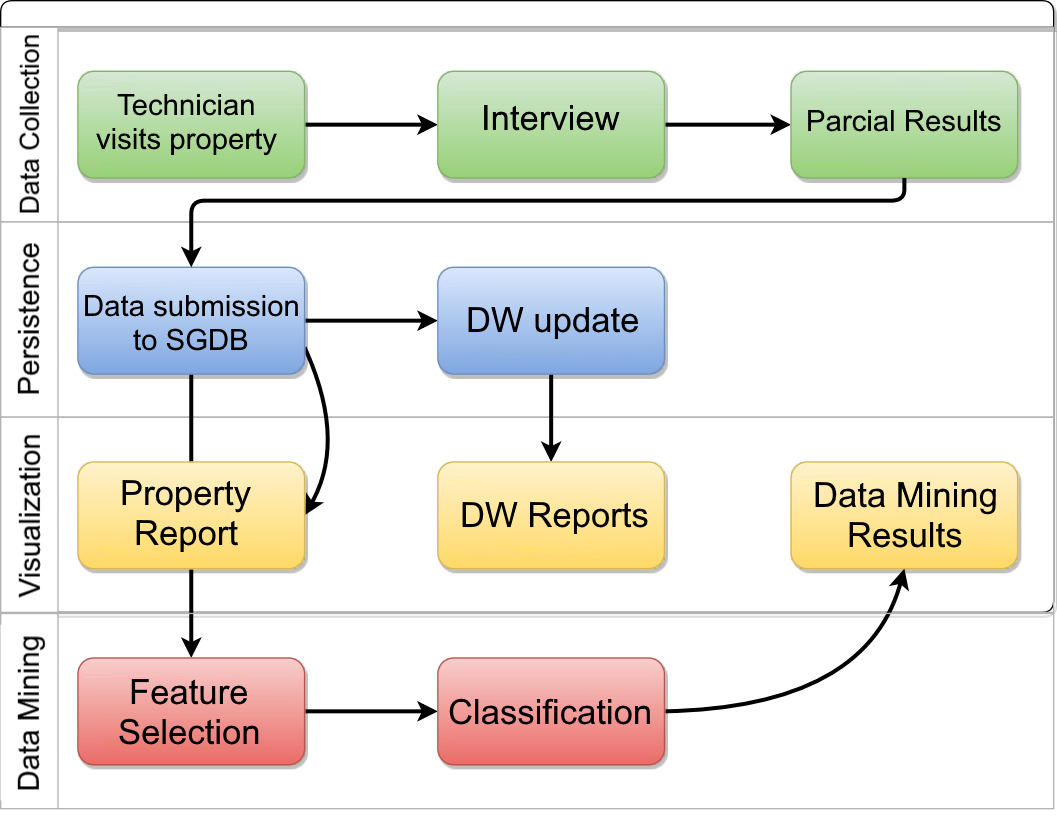}
    \caption{\isadigital's steps of data processing}
    \label{fig:isaProcess1}
\end{figure}

Figure~\ref{fig:isaProcess1} shows the process stream, beginning with the appliance of the ISA Questionnaire through the \isadigital~desktop client, and ending with the data mining techniques processing on the collected questionnaires, to aid and generate new information to be visualized in the web application. In the first level, colored in green, the steps for the appliance of ISA are presented. Initially, an agrarian technician visits a participating property, to start the appliance of the ISA Methodology. On this visit, a person (owner, manager, specialist or other with knowledge regarding the rural property) is interviewed and aids the technician with the fulfilment of the ISA questionnaire, a partial result over the collected data is generated and available in the data collection client. Data storage and management are presented in the second level. In the third level, yellow, the data visualization steps are shown. There are included: the reports generated for each questionnaire submitted, data warehouse and data mining reports processed and generated over aggregations of those reports, all accessible through different screens and sections of the web application. In the last level, red, the steps to execute the data mining techniques (involving machine learning) are displayed.

\subsection{Modules}\label{sec:app:modules}

In this section, we describe some user interfaces from different modules of the \isadigital~ system. We present and describe the data collection client module, the web application for data visualization and management module, and the Data Warehouse module.

\subsection{Data collection client}\label{sec:app:cliente}

\begin{figure}[!htb]
    \centering
    \includegraphics[width=0.85\linewidth]{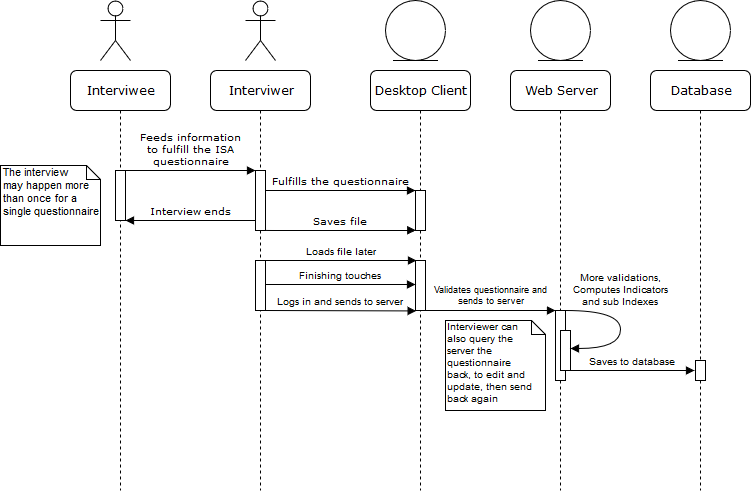}
    \caption{The process of interviewing and sending/editing a questionnaire for a rural property in \isadigital, through the desktop client}
    \label{fig:umlClient}
\end{figure} 

The information and data collected in the first step of the process are done by the fulfilment of the ISA questionnaire, divided into tabs and forms, presented, managed and validated by the JavaFX (for Java 8) desktop client application, written in Java 8 for the Java Runtine Environment, supporting Windows, various Linux distributions and MacOS, as long as the user has Java 8u121 (or a more recent version) installed. The application contemplates all the data listed in the appendix \ref{ch:appendix}.

The data collection client is responsible for formatting, validating and applying type standardization overall data inputted by the technician, so it is stored consistently and is acceptable by the online persistence layer later. The client also provides the partial results computation of a ISA questionnaire on the go.

The forms are made with text areas, text fields, number fields, tables of numbers and text, check boxes and combo boxes. The forms are divided by tabs (representing a section of the ISA Methodology), the tabs are divided by 3 main questionnaires - [General] Questionnaire, Geoprocessing, and Indicators -, switchable in a lateral left menu (those replace the Excel sheet tabs). The sum of the three forms composes the final questionnaire. The data input controls were projected as such to reduce errors and optimize the filling in overall time, as the questionnaire is lengthy, with various kinds of fields for data of different nature. In grey boxes are displayed the results of mathematical functions of the ISA Methodology, regarding the nearby text fields.

Through the desktop application, it is possible to attach images, to the questionnaire, displaying the contour and borders of the property, and its locations. That is done to highlight geographical or structural features of the properties judged relevant by the Interviewer or Interviewee.

The interviewer can collect all the required data offline, save the questionnaire to disk (in the format of a file .isa, generated by the application, on demand), and load it to edit or send it later, through the desktop application, to the online module of the \isadigital~system, upon the completion of the questionnaire or when an Internet connection is available. The questionnaire's data can also be updated after it is sent, in the case of a mistake or inconsistency, or other reasons the technician sees fit.

\subsection{Web application}

\begin{figure}[!htb]
    \centering
    \includegraphics[width=0.85\linewidth]{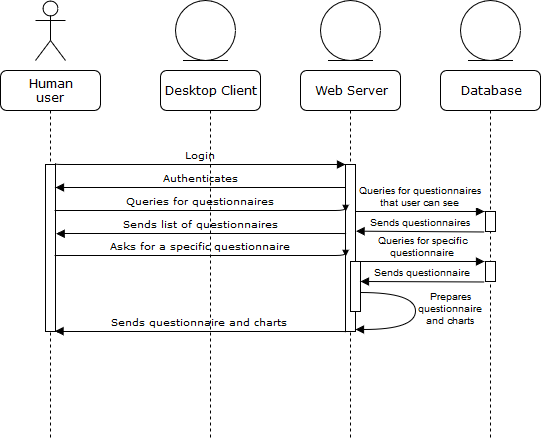}
    \caption{The process any human actor executes to query a questionnaire in the \isadigital~ system, the web component provides this service to any human user.}
    \label{fig:umlQuery}
\end{figure} 

For all users, the Web Application lists the reports the user can see, either on a table or on a map, plotting a point to the properties' coordinates (using Google Maps and it's Markers). The full ISA compliant report, enriched with interactive charts, is displayed when selected. The charts on the report plot the same information as the ones in the end of a typical ISA questionnaire Excel sheet. A report's page also offers a print service that generates a printable PDF with the contents of the page. All the users also can download the Java desktop client application through a "Downloads" web page, accessible through the navigation bar, with instructions on how to install both the client and Java, for each target operating system (Windows, Linux and MacOS).

The users that have access to the main dashboard land on it when they log in the system. The main dashboard shows a synthesis of the reports of a set of properties that can be generated by filtering questionnaires for different criteria (by project, date of collection, municipality or water basin in which the property is located, the primary source of income of a property, among other criteria). It displays the full report of the ISA Methodology, complemented with Radial Charts, but by averaging the indicators and sub-indexes and summing up monetary and unit quantities of the filtered properties' unitary reports. It also displays a Box Plot of the general balance for the properties in the filtered set, containing their sub-indexes.

The charts and tables of the main dashboard, Analytical Intelligence and Data Warehouse are accessible to Managers, aiming to offer a general visualization of the results obtained by the development of a \isadigital~ project, grouping data of participating properties and allowing for aggregations and grouping the user wishes to make.

\subsection{Analytical Intelligence}\label{sec:app:analitica}

\begin{figure}[!hb]
    \centering
    \includegraphics[width=0.85\linewidth]{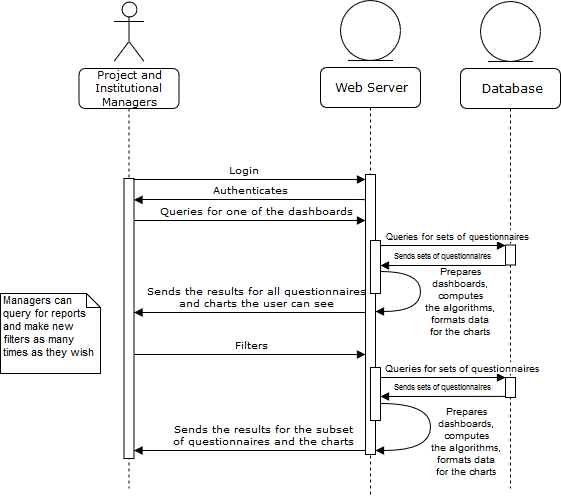}
    \caption{A high level representation of the process the human actors called Managers orchestrate in order to get aggregations of rural properties, visualizations and information from those in their dashboards.}
    \label{fig:umlIntelligence}
\end{figure} 

Analytical Intelligence methods are employed to generate more complex and detailed charts, aimed to aid Managers in deciding what actions to take to increase their properties' Sustainability Indexes further. The data (questionnaires) fed to those methods can be filtered by the itens in table \ref{tab:filteritems}. Only managers can access the panels generated by such methods.

\begin{table}[!hbt]
  \caption{Filter criterias available for the user on the Analytical Intelligence screen of \isadigital}
  \label{tab:filteritems}
  \begin{tabular}{l}
    \toprule
    Criteria \\
    \midrule
    Project (on the questionnaire's header) \\
    Year (on the questionnaire's header) \\
    Main income (Product or Service name) \\
    SENAR Region of the property \\
    Coffee Producing Region of the property \\
    State of the Union \\
    Meso Region \\
    Micro Region \\
    Municipalities \\
    Water Basin(s) the property is in \\
    \bottomrule
  \end{tabular}
\end{table}

The Analytical Intelligence techniques present in this section of the system for the user to interact with are JRip (shown through a Sankey diagram) and CFS (displayed through a simple bar chart). The other charts and graphs available are Word Cloud (of most common words filled into the free text forms of the questionnaires), Areas Trees and Scatter Chart for any one of the 21 Indicators - as in table \ref{tab:subindexesindicators} - the user wants to see from the filtered properties (or whole) set.

Finding association of attributes rules that imply satisfactory Sustainability Indexes can also help the identification of good practices to be implemented on properties currently presenting unsatisfactory Indexes. Through the JRip Sankey Diagram, one can see trends of attributes' values leading to similar \IS~ in the filtered data set. The drawing shows which attributes are associated for each rule, how many properties presented that trend, and each rule's relevancy for a bad, average or good~\IS. An example can be seen in figure \ref{fig:JRIP}.

The CFS chart allows the user to see which attributes are most relevant or more expressive to the final result for the~\IS for a questionnaire. The technique employed to create that chart takes into account all the questionnaires currently available on the system's data set at the access moment. That attribute selection technique is called \textit{Correlation Features Selection}. An example can be seen in figure \ref{fig:CFS}.

The Area Chart allows the comparison between the geographical area of each rural property with their area proportional to some indicator or sub-index which is chosen by the user.

The Scatter Chart allows the user to plot any Indicator or Sub Index as the X or Y axis of the chart, and identify relations between them on the filtered data set.

A high level representation of the process Managers orchestrate in order to get aggregations of rural properties, visualizations and information from those in their dashboards is shown in figure \ref{fig:umlIntelligence}.

\subsection{Data warehouse}\label{sec:app:dw}

A data warehouse module implemented with Pentaho Community Edition was developed and integrated into the system to give the users the capacity to compare the performance of rural properties by geographical regions and years.

The user interacts with the warehouse functions of Pentaho through a dashboard, featured of filters and bar charts. The filters appear first, allowing the user to filter properties by multiple selections of municipalities and years. Once the user sets up and confirms the filters, the dashboard fetches the required information from Pentaho.

The first bar chart displays all the Indicators\ref{tab:subindexesindicators} and the Sustainability Index averaged and grouped by the municipalities filtered, putting the aggregations over municipalities side by side for comparison. In a similar manner, it also features a Sub-Indexes\ref{tab:subindexesindicators} bar chart, grouped and averaged by municipality as well. 

The third chart is an Incomes bar chart, displaying separately the sums of \textit{Questionnaire 11.1 - Estimated annual gross income} of each property, \textit{Questionnaire 12.1 - Facilities and other betterments} inside the properties, their \textit{Questionnaire 12.1 - Machinery and Equipment} values, \textit{Questionnaire 12.3 - Animals} expenses, \textit{Questionnaire 13.3 - Total estimated value of the rural property} and \textit{Questionnaire 12.4 - Irrigation} expenses for every property in each municipality. Each municipality is represented by a group of bar charts, composed by one bar for one of the cited items. Those groups of bar charts are displayed next to each other, for comparison purposes. Those items come from the economic section of the questionnaire.

\subsection{Report and adequation plan}

\begin{figure}[!htb]
    \centering
    \includegraphics[width=0.85\linewidth]{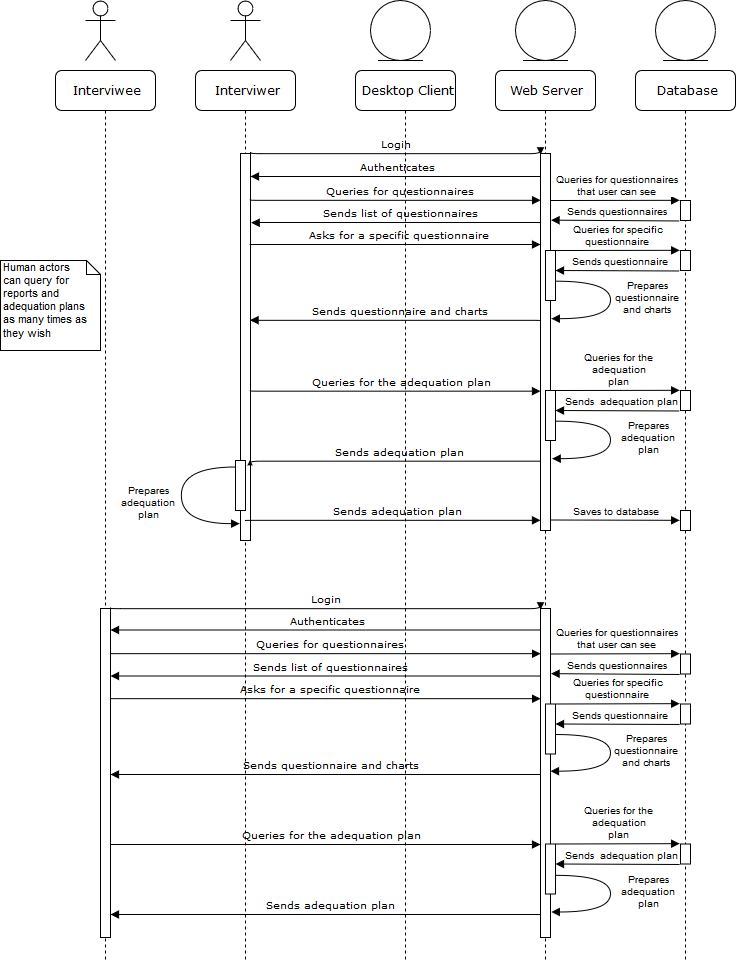}
    \caption{The process technicians and representatives of rural properties execute to use the Adequation Plan, through \isadigital}
    \label{fig:umlAdequationPlan}
\end{figure} 

For each data collection of a property in a year that a technician sends to the system, in the name of a project, a report is computed and generated with the data filled in the questionnaire. An user of the system can obtain and read those reports depending on their level of access (interviewee can visualize only the reports for properties they are responsible for; the interviewers can see all the properties' reports for all the questionnaires they submitted to the system; the managers can see reports for every property related to the projects or institutions they manage).

The ~\isadigital\ system also has a web interface implementing a module of the original ISA model called Adequation Plan For a Rural Property. This module is used by the technicians to guide the farmers on which actions to take to make their rural properties more sustainable, according to the ISA methodology. There is a text area for suggestions regarding each indicator. The technician fills those forms with the required actions for the farmer to implement. The results should be observed and evaluated by the technician in future visits to each rural property.

An UML diagram featuring the interaction of actors since logging in \isadigital~to write and read an adequation plan is shown in figure \ref{fig:umlAdequationPlan}.
\section{Cases}

\label{ch:case}

In this section, we present an experimental case study using the data set shown in the methodology section.
In \secstr~\ref{sec:case:data set}, the characterization of the data set is explained. These analyses are relevant for all further discussion presented in this paper.

\subsection{data set}\label{sec:case:data set}
The \textit{Embrapa Pecuária Sudeste} was responsible for the Balde Cheio project to foster innovations to increase the profits of dairy producers~\citep{borges2011estudo}. It was held in Minas Gerais by the FAEMG system, whose make efforts to spread the program for all regions of the state. The program consists of providing training of technicians contracted by partner entities and testing new technologies to assist producers of milk. The new technologies are also monitored for their environmental, economic and social impact .

This project carries out technology transfer for milk producers and related entities. The results of this process are the development of the sector and increase the profitability of the rural producer. With the increase of profitability of the property, it becomes more viable the permanence of the workers involved in the milk production in the field, instead of looking for opportunities in the city~\citep{novo2013feasibility}.

The data used in this work were collected by the project Balde Cheio, which was applied by FAEMG, using the system presented in the former section, \isadigital. In another words, \isadigital~ was one of the technologies tested by Balde Cheio in 2016.

\begin{figure}[htbp]
     \centering
     \includegraphics[width=1.00\textwidth]{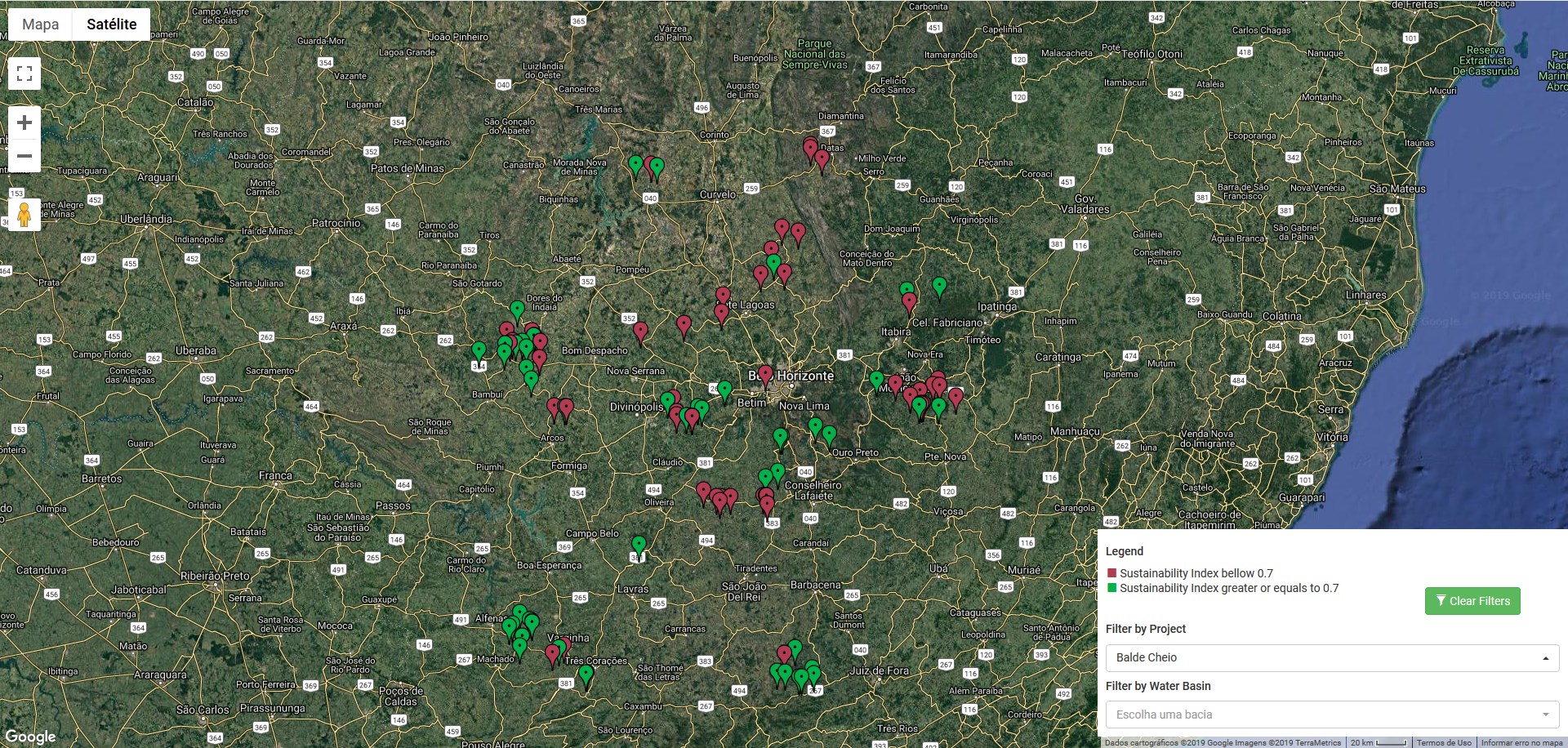}
     \caption{\isadigital~: Geographical Visualization Chart, displaying participating properties in \isadigital~ where Balde Cheio is being applied.}
     \label{fig:baldeCheioCities}
\end{figure}

\figstr~\ref{fig:baldeCheioCities} presents the properties of the state of Minas Gerais where the Balde Cheio Program has been applied and had the \isadigital~questionnaire sent to our system. The map was generated by \isadigital~, using Google Maps to plot the markers. Green markers mean the Sustainability Index for that property is greater or equals than 0.7, red markers mean it is below 0.7. In total there are 317 municipalities and 1929 participating properties in Balde Cheio, the number of properties both in Balde Cheio and \isadigital~, however, is 100. The maximum number of participating properties in a municipality is 112. Of the 317 municipalities, 75\% of them have up to 7 properties participating in the program. In our system, there are 100 questionnaires, for 100 unique properties.

FAEMG collected data for \isadigital~from 100 rural properties enlisted in Balde Cheio, having their technicians visit the property and interview a person responsible for it, meeting the ISA Methodology requirements. The properties are scattered across several sub-basins as shown in \tabstr~\ref{sec:case:basins}. The collection was carried out between August 2016, and January 2017 in a group of properties and municipalities defined by FAEMG.

\begin{table}
    \centering
    \begin{tabular}{ r  l  }  
    \toprule
    \textbf{Prop. Count} & \textbf{Sub-basin} \\ \midrule
    19	& Tributaries do Alto São Francisco \\
    16	& Basin of Paraopeba River \\
    13	& Basin of Pará River \\
    11	& Basin of Das Velhas River \\
    9	& Basin of Piracicaba River \\
    8	& Basin of Furnas Reservoir \\
    5	& Basin of Verde River \\
    4	& Basin of Alto Rio Grande \\
    4	& Basin of tributaries from Minas Gerais of the rivers Preto and Paraibuna \\
    4	& Basin of Piranga River \\
    3	& Basin of Santo Antônio River \\
    3	& Basin of waters around the dam of Três Marias \\
    1	& Hydrographic basin of São Francisco River \\ 
    \bottomrule
    \end{tabular}
    \caption{Hydrographic Sub-basins}
    \label{sec:case:basins}
\end{table}

\begin{figure}[htbp]
    \centering
    \includegraphics[width=.6\textwidth]{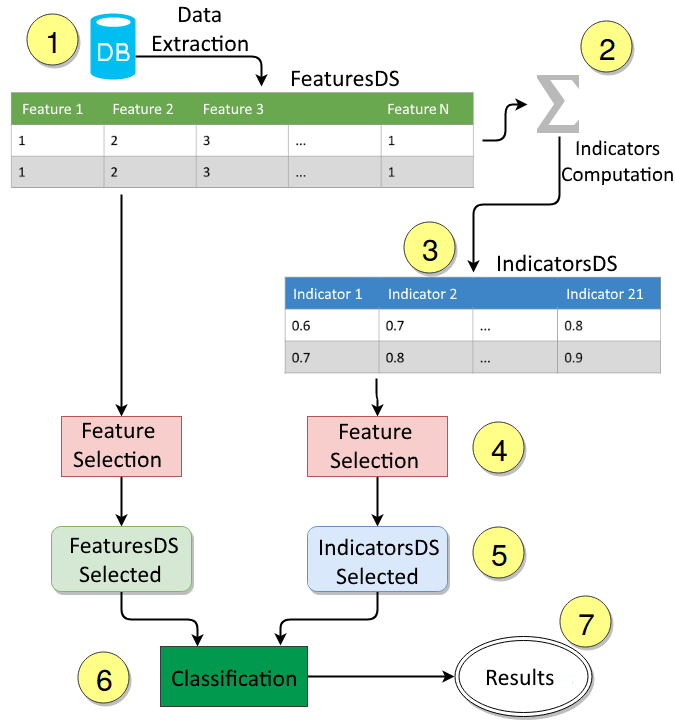}
    \caption{Methodology}
    \label{fig:methodology}
\end{figure} 

The complete list of available variables is in Appendix~\ref{ch:appendix}, our model uses 87 of those. \figstr~\ref{fig:methodology} summarizes the methodology of this work: The first step consists of formatting the attributes from the ISA questionnaires data, the attributes are mapped to features and that set with data from all the participating questionnaires is called "FeaturesDS"; During the second step, the 21 sustainability indicators for each questionnaire are calculated; In the third step, the data set "IndicatorsDS" is generated by formatting in the 21 sustainability indicators computed for each entry of "FeaturesDS"; For the fourth step, \textit{Features Selection} is applied separately to both "FeaturesDS" and "IndicatorsDS"; The fifth step consists of generating reduced versions of both data sets, filtering out the attributes that did not get picked by the \textit{Features Selection} step; For step six, the classification algorithms are executed with both data sets; the last and seventh step consists of analysis of the results.

\begin{figure}[htbp]
    \centering
    \includegraphics[width=0.6\textwidth]{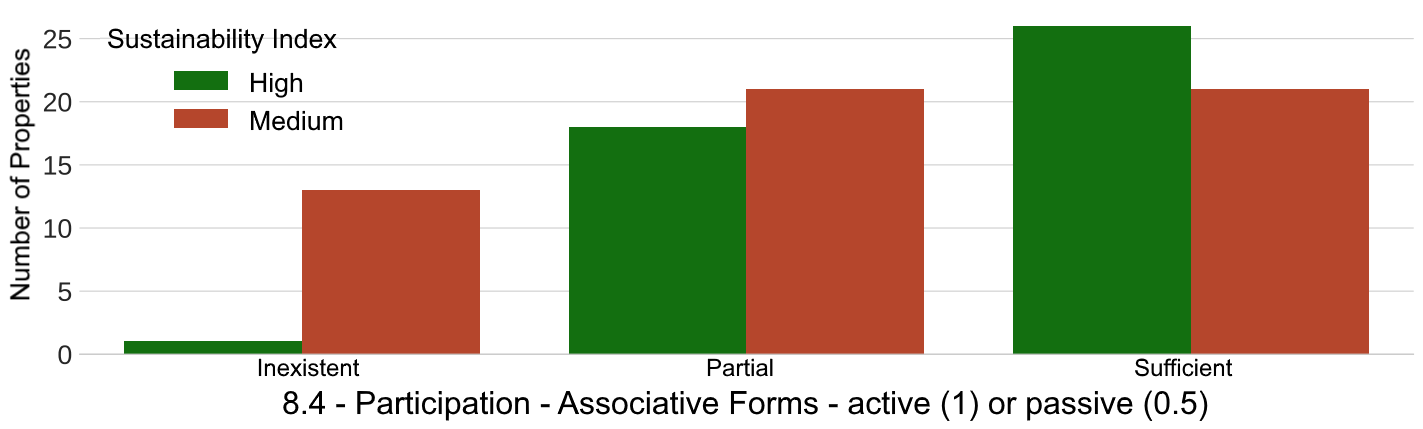}
    \caption{Categorical Variable: Associative Forms}
    \label{fig:associativeForms}
\end{figure}

In \figstr~\ref{fig:associativeForms} we can see how much the producer is involved in activities and their level of sustainability. Whenever there is no form of association, there is a lower incidence of properties with adequate IS. We also observe that with the increase of the \textit{Association Forms} level in the properties there is also an increase of \IS. It may indicate that this is a variable that influences the sustainability of properties.
  
\begin{figure}[htbp]
    \centering
    \includegraphics[width=0.6\textwidth]{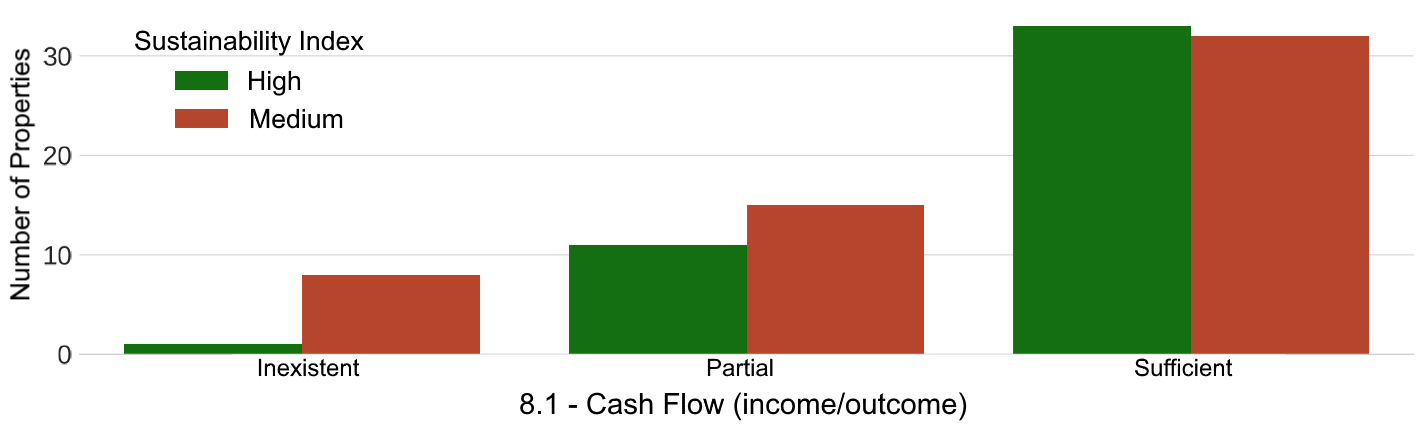}
    \caption{Categorical Variable: Cashflow}
    \label{fig:cashFlow}
\end{figure}

The variable \textit{8.1 - Cash Flow} identifies whether the producer has control of rural property expenses and receipts. In \figstr~\ref{fig:cashFlow} it is possible to observe that more than 60 properties have sufficient control of their cash flow.

\begin{figure}[htbp]
    \centering
    \includegraphics[width=0.6\textwidth]{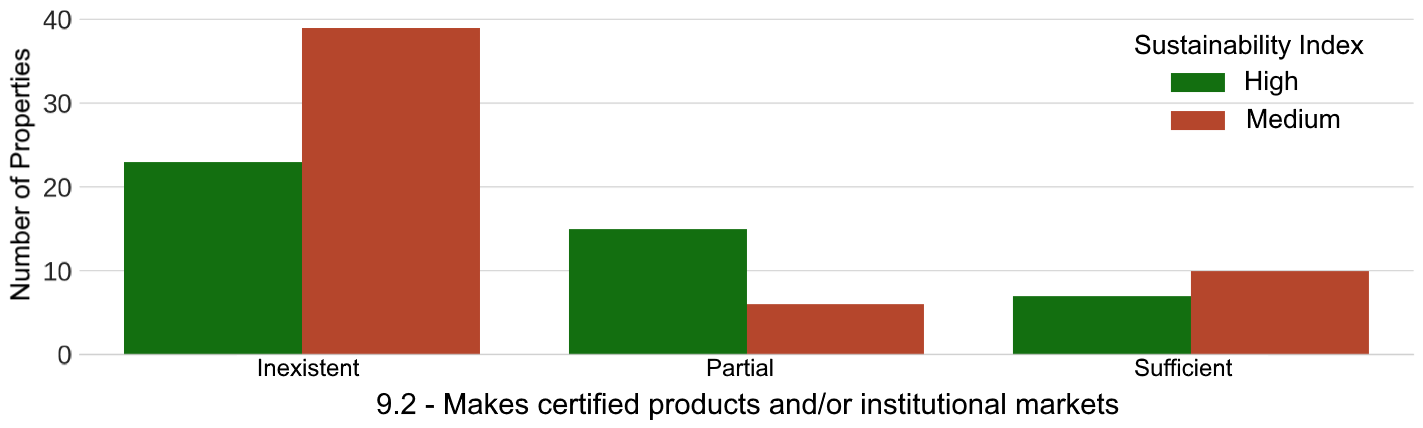}
    \caption{Categorical Variable: Certified Products}
    \label{fig:certifiedProducts}
\end{figure}

The variable \textit{9.2 Certified Products} shows whether the property produces certified products. As seen in \figstr~\ref{fig:certifiedProducts} more than 50\% of the properties do not have any certification for their products.

\begin{figure}[htbp]
    \centering        
    \includegraphics[width=0.6\textwidth]{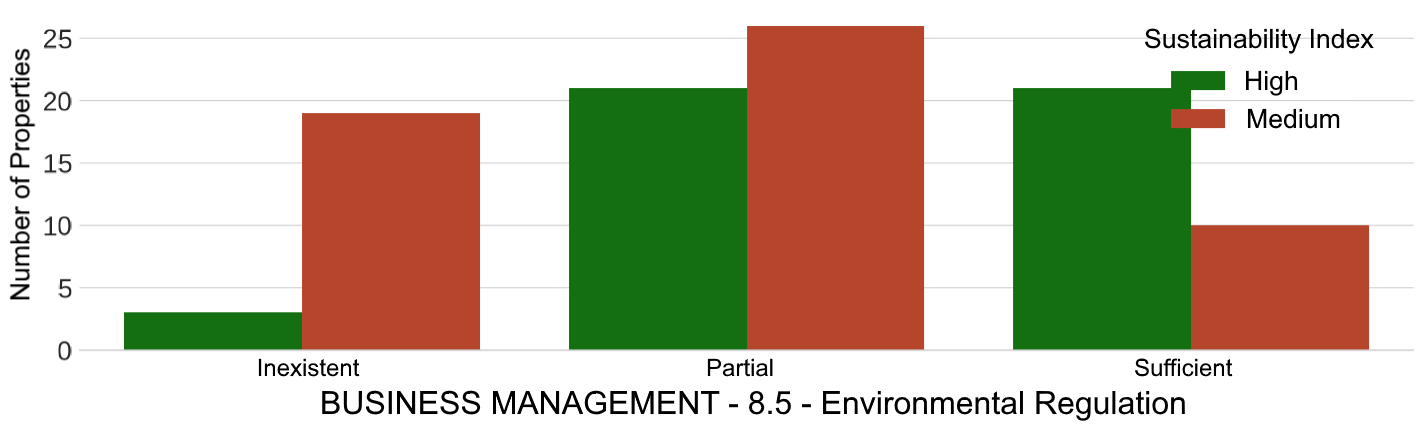}
    \caption{Categorical Variable: Environmental Regulation}
    \label{fig:environmentalRegulation}
\end{figure}

The variable \textit{8.5 - Environmental Regulation} verifies the Environmental Regulation of the producer concerning water, environmental license, and regularization of the Legal Reserve (RL) and Permanent Preservation Areas (APP). \figstr~\ref{fig:environmentalRegulation} shows that less than 25\% of the properties did not have any regularization.

\begin{figure}[htbp]
    \centering        
    \includegraphics[width=0.6\textwidth]{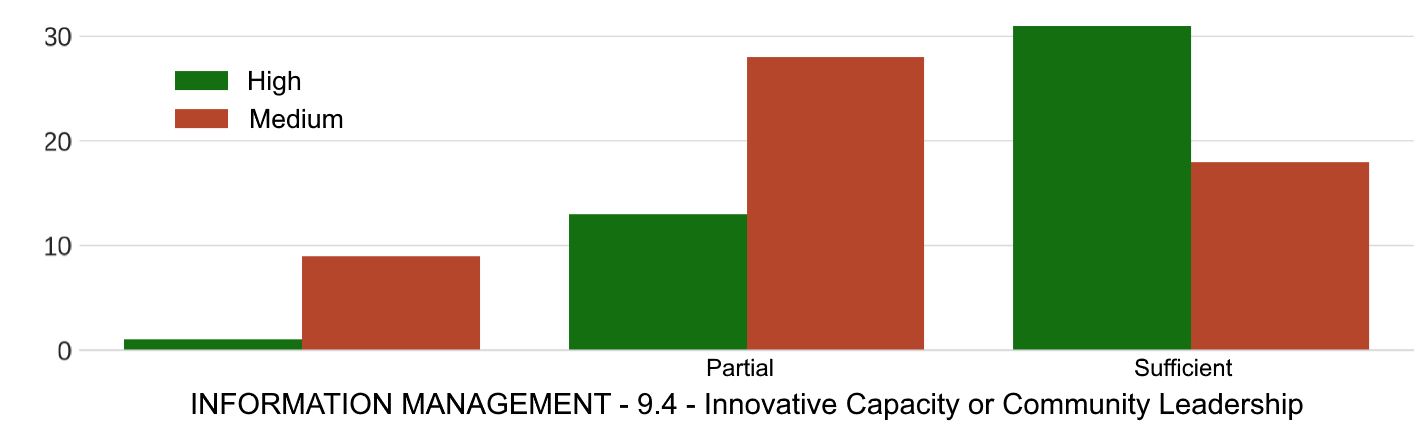}
    \caption{Categorical Variable: Capacity for Innovation and Leadership}
    \label{fig:innovationCapacity}
\end{figure}

The variable \textit{9.4 - Capacity for Innovation and Leadership} measures the producer level of innovating in the practices that are adopted on his property as well as whether he has a leadership role in his community. \figstr~\ref{fig:innovationCapacity} presents the charts for this variable related to Sustainability. Among the three levels of \textit {Capacity for Innovation and Leadership} the sufficient level is the one that contains the highest number of properties at that level more than 30 \% of the properties have above a satisfactory \IS. For the group with \textit{Partial Innovation and Leadership} there are fewer properties with satisfactory \IS than average. It is necessary sufficient capacity for innovation and leadership for long-term investments such as staff training, courses, and adoption of new practices. This factor may indicate why properties that are not already presenting satisfactory \IS~ are the minority in the group of properties with partial innovation.

%

\begin{figure}[htbp]
    \centering
    \includegraphics[width=1\textwidth]{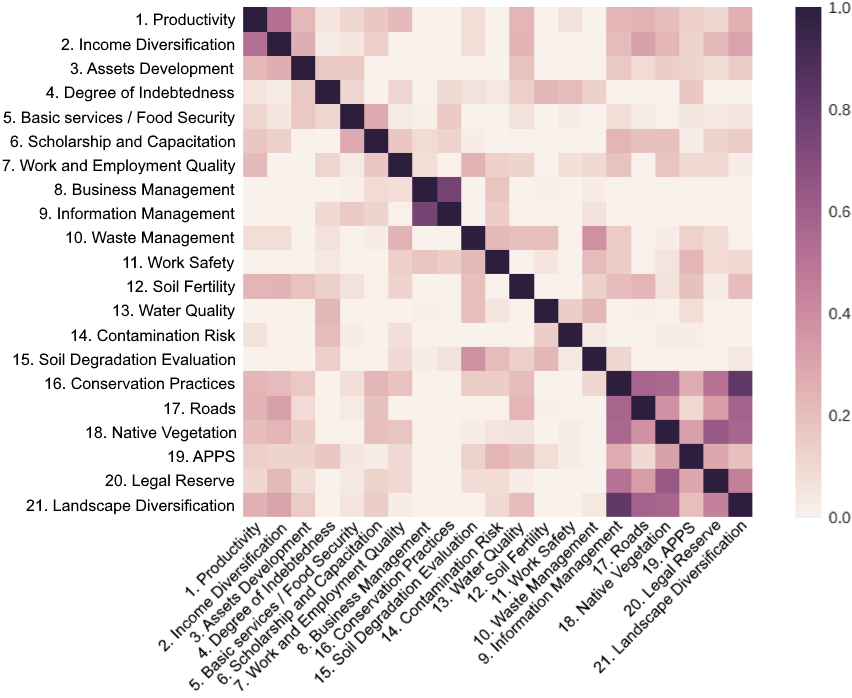}
    \caption{Indicators Heatmap}
    \label{fig:heatMapAll}
\end{figure}

In \figstr~\ref{fig:heatMapAll} we can see correlations between the sustainability indicators. Some groups of variables stand out because they present high correlation, such as the following groups:
\begin{itemize}
 \item Income Diversification and Productivity
 \item Information Management and Property Management
 \item Conservation Practices, Roads, Native Vegetation, APPs, Legal Reserve and Landscape Diversification.
\end{itemize}

Using the general dashboard available in \isadigital\, we get the average Indicator and Sub-Indexes scores for the set of 100 properties participating in the Balde Cheio project:

\begin{figure}[!htbp]
    \centering        
    \includegraphics[width=0.7\textwidth]{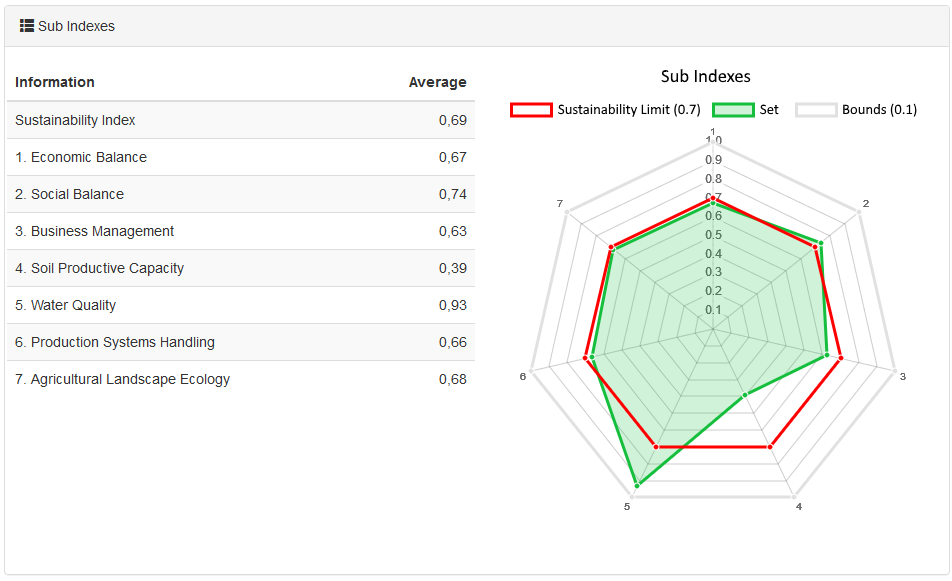}
    \caption{\isadigital~: Averages Radar Chart of the Sub-Indexes for the 100 properties participating in both \isadigital~ and Balde Cheio.}
    \label{fig:baldeCheioSubIndexes}
\end{figure}

In \figstr~\ref{fig:baldeCheioSubIndexes}, we can see that this set of properties averages badly regarding \textit{Soil Productive Capacity} in the Sub-Indexes, which is explainable by the fact those are milk-producing properties. Thus they are enrolled in a program for milk producers (Balde Cheio). The 100 properties of Balde Cheio tend to exceed in the \textit{Water Quality Indicator}, and as the chart shows, their average is slightly below the recommended mark when it comes to \textit{Production Systems Handling}, \textit{Business Management} and \textit{Economic Balance}. The other Sub-Indexes are very close to the desired line, on average.

\begin{figure}[!htbp]
    \centering        
    \includegraphics[width=0.7\textwidth]{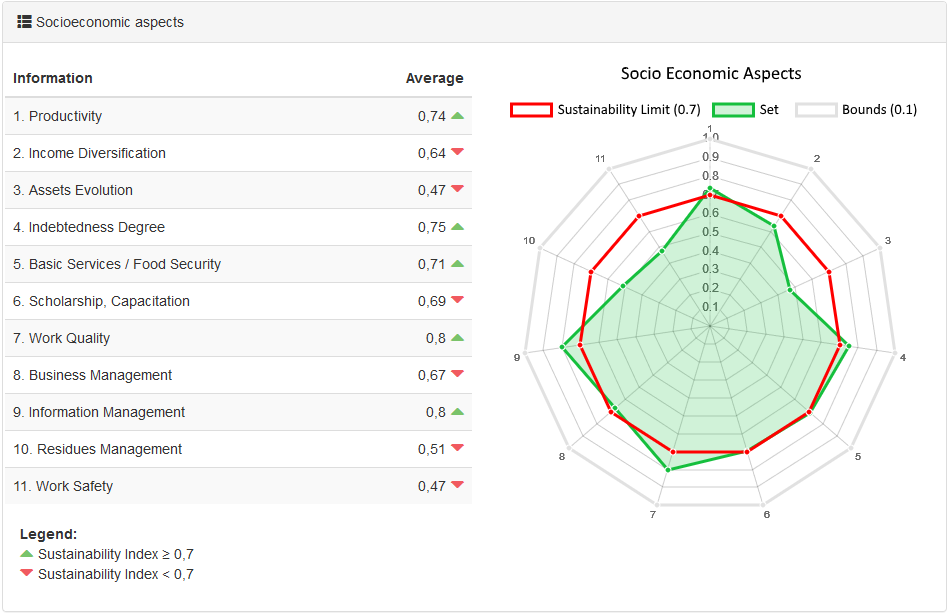}
    \caption{\isadigital: Averages Radar Chart of the first 11 Indicators for the 100 properties participating in both \isadigital~ and Balde Cheio.}
    \label{fig:baldeCheioSocioAspects}
\end{figure}

\figstr~\ref{fig:baldeCheioSocioAspects} presents the Socioeconomic Indicators. The system show us that the properties of this set present below desirable averages for \textit{3. Assets Development}, \textit{10. Residues Management} and \textit{11. Work Safety}. \textit{3. Assets Development} is an indicator of the Sub-Index \textit{Economic Balance} and the other two indicators are aggregated in the \textit{Business Management} sub-index. It's possible to see that those 3 indicators impact the results expressed in \figstr~\ref{fig:baldeCheioSubIndexes}.

\begin{figure}[!htbp]
    \centering        
    \includegraphics[width=0.7\textwidth]{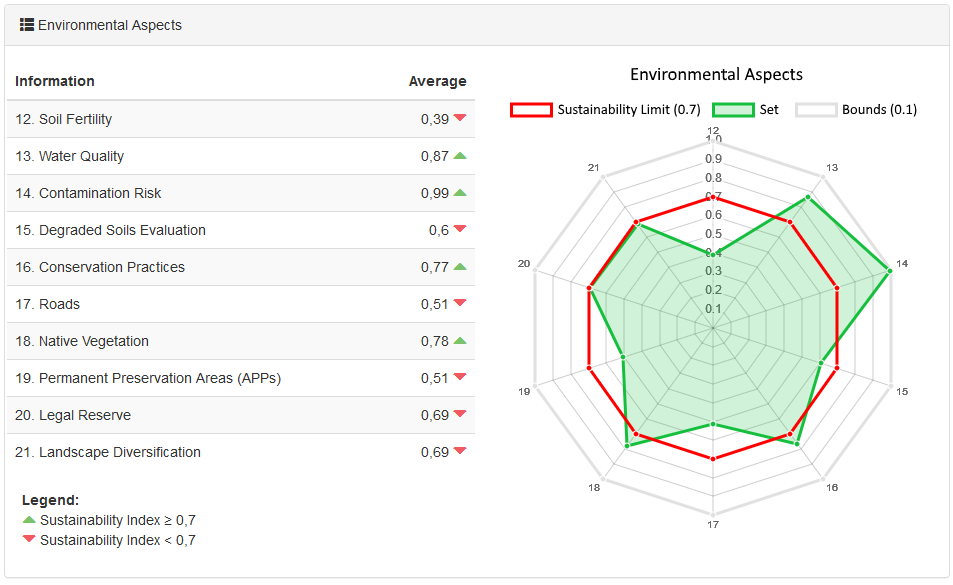}
    \caption{\isadigital~: Averages Radar Chart of the Environmental Aspects related Indicators for the 100 properties participating in both \isadigital~ and Balde Cheio.}
    \label{fig:baldeCheioEnvironmentalAspects}
\end{figure}

For the environmental Indicators, displayed in \figstr~ \ref{fig:baldeCheioEnvironmentalAspects}, Balde Cheio's properties do generally well in \textit{14. Water Contamination Risk (containment)} - a part of the \textit{Water Quality} sub-index, while also generally being located in areas of hard access and which presents, on average, a bad soil for farming - as the Indicators \textit{17. Roads} (part of \textit{Handling of the Production Systems}) and \textit{12. Soil Fertility} show. On average, for those properties, the \textit{19. Permanent Preservation Areas} (an indicator weighted in the \textit{Ecology of the Rural Landscape} sub-index) score is low, meaning the areas reserved for preservation of the native vegetation are either below the recommended in size or are in bad conservation state.

\begin{figure}[!htbp]
    \centering        
    \includegraphics[width=0.8\textwidth]{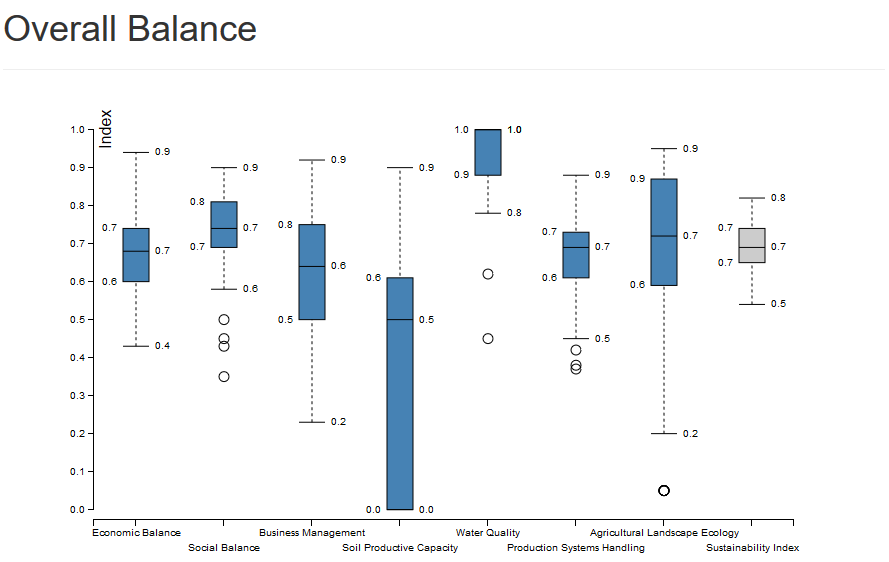}
    \caption{\isadigital~: Averages Radar Chart of the Environmental Aspects related Indicators for the 100 properties participating in both \isadigital~ and Balde Cheio.}
    \label{fig:baldeCheioOverall}
\end{figure}

The dashboard also displays a box plot of the Sub-Indexes in the shape of blue bars presenting the final Sustainability Index in the last column, represented by a grey bar.

As pointed out by the radar charts, the \textit{Soil Productive Capacity} indicator performed the worse for those milk producing properties, the majority of properties are below the healthy (0.7) line for \textit{Business Management}, \textit{Production Systems Handling} and \textit{Economic Balance} as well. \textit{Water Quality} is very high overall. As the first radar chart points out, the set has most of its properties above or near the minimum desirable Sustainability Index.

It is worth mentioning that the main dashboard panel is only capable of generating averages or sums for the sets it receives. The Analytical Intelligence panel, on the other hand, can show us both the results of JRip and CFS for that subset.
\begin{figure}[!htbp]
    \centering        
    \includegraphics[width=0.6\textwidth]{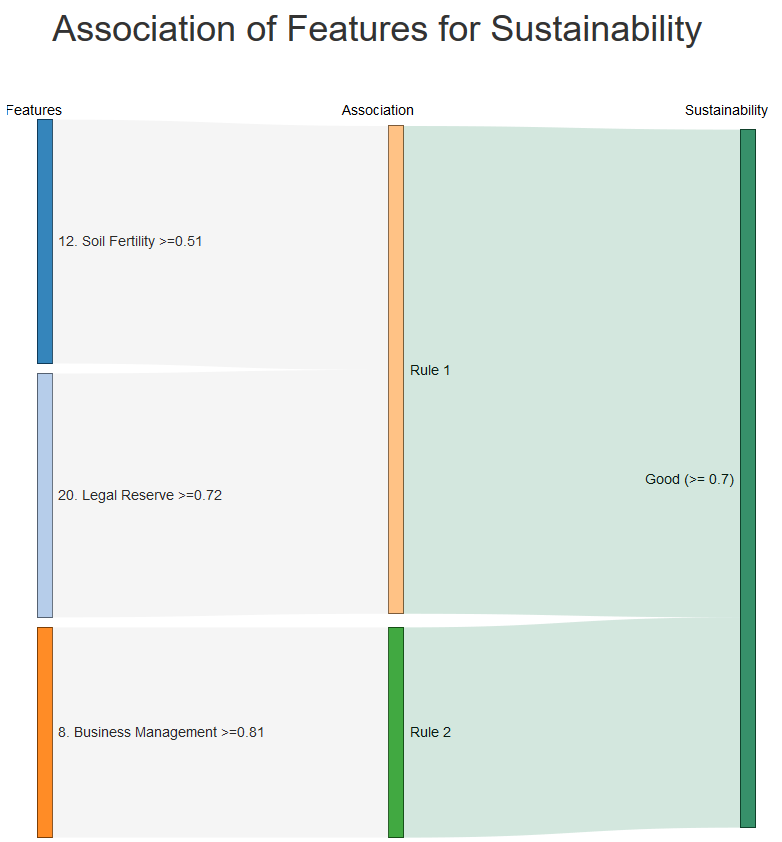}
    \caption{\isadigital~: Averages Radar Chart of the Environmental Aspects related Indicators for the 100 properties participating in both \isadigital~ and Balde Cheio.}
    \label{fig:JRIP}
\end{figure}

The JRip Sankey Diagram (\figstr~\ref{fig:JRIP}) generated by \isadigital~ for that Balde Cheio's set of properties shows that if the property has a high \textit{Business Management} Indicator it is highly likely to be ranked with a high Sustainability Index. As shown in the box plot and the radar charts for averages, the majority of those properties score below recommended for the \textit{Business Management} Indicator.

The associative rules found by JRip (\figstr~\ref{fig:JRIP})  for our data set shows that those - 25 properties - which managed to score greater or equal the recommended value ($0.70$) for \textit{Business Management} are highly likely to also score an above recommended Sustainability Index. As the Sustainability Index is the average of all the Indicators, those properties also tend to do well for the other Indicators as well.

This Figure shows the most critical rules associations that lead to a good sustainability index (SI) discovered by JRIP. A good SI is characterised by a score equals to or greater than $0.70$. For this dataset, the combination of good ratings on Indicators 12, 20 and 8 are characteristic of the majority of properties that also present a good SI. The association of high scores for Indicators 12 and 20, the first rule displayed in the graph, is covered by 58\% of the rural properties with a good SI score. The second rule in the graph, which is having a \textit{Business Management} score equals to or greater than 0.81, has coverage of 50\%. The intersection of these rules shows us the effectiveness of the \textit{8.1 Access to Technical Assistance} provided to the agricultural property manager: with proper assistance of a technician, the producer can better understand the soil quality of the property, the best suitable culture to plant on it and also how to treat or deal correctly with any imperfections on the soil, all affecting the 12th Indicator, \textit{Soil Productive Capacity}. \textit{8. Business Management} is also important to ensure the financial viability of the entrepreneurship. A good compliance to \textit{20. Adequacy of Permanent Preservation Areas} by rural properties is also essential for the preservation of the lands and water resources.

The \figstr~\ref{fig:JRIP} that for properties that got a \textit{Soil Fertility} Indicator score near the median of the set ($0.50$) and simultaneously scored a \textit{Legal Reserve} slightly higher than the recommended value ($0.70$) also tend to present a satisfactory \IS.

\begin{figure}[!htbp]
    \centering        
    \includegraphics[width=0.8\textwidth]{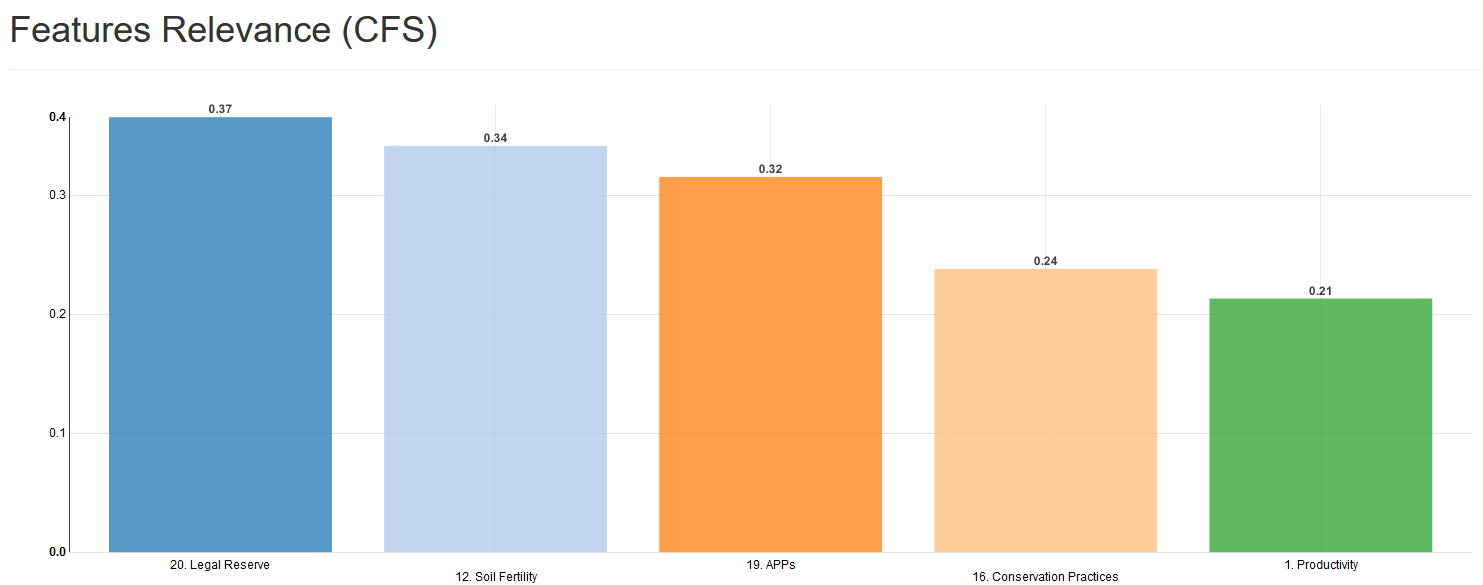}
    \caption{\isadigital~: Averages Radar Chart of the Environmental Aspects related Indicators for the 100 properties participating in both \isadigital~ and Balde Cheio.}
    \label{fig:CFS}
\end{figure}

The CFS chart generated by \isadigital~ for that subset, in \figstr\ref{fig:CFS}, shows that such algorithm selected the indicators \textit{20. Legal Reserve}, \textit{12. Soil Fertility}, \textit{19. APPs}, \textit{16. Conservation Practises} and \textit{1. Productivity} as the most relevant Indicators that weighted in the \IS~ of a property.
\textit{20. Legal Reserve} and \textit{19. APPs} are a part of the \textit{Ecology of the Rural Landscape} sub-index, \textit{12. Soil fertility} has a Sub-Index of it's own called \textit{Soil Productive Capacity}, \textit{16. Conservation practices} is one of the factors for the \textit{Handling of the Production systems} Sub-Indexes and \textit{1. Economic Balance} is partially composed by \textit{1. Productivity}.


\section{Results}\label{sec:case:experiments}

%

%

In this section, the experiments and the processes that led to the results presented so far are detailed. The results detailed here were obtained by using the Weka software ~\citep{witten2016data}.

With the database initially collected two data sets were sorted and grouped. The first set \atributosDS~is made of 87 attributes; meanwhile, the second one, \indicadoresDS, contains the 21 indicators described in the ISA Model plus the Sustainability Index (SI). In both data sets, the SI was categorized in three levels: Low for when the SI is between the interval $[0; 0,5]$, Medium when it's between $[0,5; 0,7]$ and High for values between $[0,7; 1,0]$. The High level is, according to the ISA Model, the interval in which the value is considered satisfactory, the other two levels being insufficient and meaning that the property needs intervention. In the collected data, all the properties had a SI evaluated as in Medium or High level, meaning we have a problem regarding the binary classification.


For each one of the two data sets, we applied the attribution selection techniques \textit{Correlation-based Feature Selection}~(CFS) and \textit{Information Gain}~(InfoGain). The hypothesis was that a smaller set of attributes could reduce the time needed for the application of the ISA Model, implying the collection of a smaller set of data of a property to know its Sustainability Index, with the same or satisfactory results when compared with the full methodology expected data.

When using CFS with the data set \atributosDS, that contains every attribute, the search method \textit{BestFirst} was deployed. This method does a greedy search on the data set checking all the possibilities. The attributes that were categorical, such as the ones listed on \secstr~\ref{sec:case:data set}, were treated as numeric, assuming the values $0$, $0,5$ and $1$ (inexistent, partial and satisfactory, respectively). For CFS on \indicadoresDS\, the same setup was deployed, excluding the categorical attributes, and all data is in the interval $[0;1]$. The validation method was cross-validation with 5 \textit{folds}. The selection found by CFS for \atributosDS is displayed in the table \ref{tab:selectionCFS}.  With \indicadoresDS, the selection found by CFS is displayed in the table \ref{tab:indicatorsonlycfs}.

\begin{table}
  \caption{Selection found by CFS for \atributosDS}
  \label{tab:selectionCFS}
  \begin{tabular}{l p{12cm}}
    \toprule
    Selection \\
    \midrule
    Questionnaire 11. & Gross Income of the business R\$/year \\
    Questionnaire 11. & Sources of the gross income of the rural business \\
    Questionnaire 14.2 & Number of natural lakes and ponds \\
    Indicators 8.5 & Environmental Regulation \\
    Indicators 9.1 & Researches information to optimize the sells of products \\
    Indicators 9.4 & Capacity to innovate or participate in leadership within the community \\
    Indicators 12.3 & Phosphorus Availability - Index \\
    Indicators 12.3 & Phosphorus Availability - Results \\
    Indicators 12.5 & Exchangeable Magnesium - Index \\
    Indicators 12.4 & Exchangeable Calcium - Index \\
    Indicators 12.7 & Active acidity (pH) - Index \\
    Indicators 12.9 & Effective CTC - Index \\
    Indicators 12.10 & Base Saturation (\%) - Index \\
    Indicators 16.2 & Level of adoption of strategies for conservation and preservation of water bodies in the rural property. \\
    \bottomrule
  \end{tabular}
\end{table}

\textit{Environmental Regulation} takes into account the use of water, Legal Reserves, Permanent Preservation Areas and proper licensing and obedience to environmental norms. The "Researches information to optimize the sells of products field" can have three values: 0, for inexistent, 0.5 for partial and 1.0 for sufficient. Part of the Information Management sub-index, that field represents the perception of how much effort the producer or other people responsible for the property put into research information to increase sales and/or attempt to diversify and reach new buyers).

Some of the attributes selected contain the suffix "Index" and "Result" in the name. Those attributes are related to the soil productive capacity of the property and are part of the computation that generates Indicator 12. The attributes with the suffix "Index" are obtained by measurement or analysis, made by a technician, using a physically collected soil sample from the property. The attributes with suffix "Result" are results from equations that receive the "Index" suffix attributes as parameters. The values with the suffix "Result" are intermediary to the computation of Indicator \textit{12. Soil Productive Capacity}.

\begin{table}
  \caption{Selection found by CFS for \indicadoresDS}
  \label{tab:indicatorsonlycfs}
  \begin{tabular}{l}
    \toprule
    Selection \\
    \midrule
    4. Degree of Indebtedness \\
    8. Business management \\
    9. Information management \\
    12. Soil Fertility \\
    16. Conservation Practices \\
    19. APPs (Permanent Preservation Areas adequation) \\
    20. Legal Reserve (Adequation of the property with the Legal Reserve standards) \\
    \bottomrule
  \end{tabular}
\end{table}

For \texttt{InfoGain} a \textit{ranking} was used to select the attributes that better define sustainability. That \textit{ranking} is constructed with individual evaluations for each attribute. Like CFS, the categorical attributes of the ~\atributosDS\ data set were treated as numeric. For \texttt{InfoGain}, the data set \indicadoresDS\ was deployed with the same configuration, except in this case there are no categorical attributes and all data is in the interval between $[0;1]$. The evaluation method was cross evaluation with 5 \textit{folds}. With the usage of \atributosDS, the selection found by \texttt{InfoGain} is featured in the table \ref{tab:selectioninfogain}. Using \indicadoresDS, the selection found by \texttt{InfoGain} is listed in the table \ref{tab:indicatorsonlyinfogain}.

\begin{table}
  \caption{Selection found by InfoGain for \atributosDS}
  \label{tab:selectioninfogain}
  \begin{tabular}{l l}
    \toprule
    Identification & Selection \\
    \midrule
    Questionnaire 11 & Sources of the gross income of the rural business \\
    Questionnaire 12.1 &  Facilities and other betterments (R\$) \\
    Indicators 8.5 & Environmental Regulation \\
    Indicators 9.3 & Adoption of innovative techniques \\
    Indicators 9.4 & Capacity to innovate or participate in leadership within the community \\
    Indicators 12.3 & Phosphorus Availability - Index \\
    Indicators 12.3 & Phosphorus Availability - Results \\
    Indicators 12.4 & Exchangeable Calcium - Results \\
    Indicators 12.5 & Exchangeable Magnesium - Index \\
    Indicators 12.5 & Exchangeable Magnesium - Results \\
    Indicators 12.7 & Active acidity (pH) - Index \\
    Indicators 12.9 & Effective CTC - Index \\
    Indicators 12.10 & Base Saturation Base Saturation (\%) - Index \\
    Indicators 16.2 & Level of adoption of strategies for conservation and preservation of water bodies in the rural property \\
    \bottomrule
  \end{tabular}
\end{table}

\begin{table}
  \caption{Selection found by InfoGain, for \indicadoresDS}
  \label{tab:indicatorsonlyinfogain}
  \begin{tabular}{l}
    \toprule
    Selection \\
    \midrule
    4. Degree of Indebtedness \\
    7. Work Quality \\
    8. Business management \\
    9. Information management \\
    12. Soil Fertility \\
    16. Conservation Practices \\
    19. APPs (Permanent Preservation Areas adequation) \\
    20. Legal Reserve (Adequation of the property with the Legal Reserve standards) \\
    \bottomrule
  \end{tabular}
\end{table}

We set up the algorithms using the default parameters, as the table \ref{tab:algorithmssetup} shows.

\begin{table}
  \caption{Algorithms setup}
  \label{tab:algorithmssetup}
  \begin{tabular}{l p{12cm}}
    \toprule
    Algorithm & Setup \\
    \midrule
	AdaBoost & 10 iterations the classifier used being \textit{Decision Stump}. \\ \hline
	Na\"ive Bayes & No additional parameter. \\ \hline
	J48 & Confidence factor of $0,25$ and the minimum number of objects per sheet was $2$. \\ \hline
	JRip & Quantity of data used for the (\textit{folds}) was $3$. \\ \hline
	MLP & learning rate was $0,3$, momentum $0,2$ and the number of hidden layers was half of the quantity of attributes summed with the number of classes.\\ \hline
	RandomForest & $100$ iterations with unlimited height.\\ \hline
	SVM & Implementation of LibSVM with classification type C-SVC, radial kernel, $\mu=0,5$, $\gamma=0$, $\epsilon=0,001$ and loss function $0,1$.\\
    \bottomrule
  \end{tabular}
\end{table}

\tabstr~\ref{tab:multicol} shows the results of the execution of all the algorithms tested. The best results of each one are highlighted in bold, the very best result being the one underlined. The metrics used to evaluate the results were precision and recall. Precision is the ratio of relevant instances to the selected ones and recall is the ratio of selected relevant instance to the total quantity of relevant instances. The technique that obtained the best precision was Random Forest, with 94\% of precision and recall.

\begin{table}[ht]
\caption{Results of executing each algorithm over the data sets \indicadoresDS\ and \atributosDS, aided by CFS or InfoGain. Random Florest (RF), Na\"ive Bayes (NB), Precision (Prec.), Recall (Rec.) and attributes (attrib).
The best precision scored by each algorithm is written in bold font. The best result is underscored.}
\centering
\begin{tabular}{llccc|cccc}
    \toprule
    \multicolumn{2}{c}{\multirow{2}{*}{\textbf{Algorithm}}} & \multicolumn{3}{c}{\textbf{Precision}} &  \multicolumn{3}{c}{\textbf{recall}}  \\
    \multicolumn{2}{c}{} & \textbf{All attr.} & \textbf{CFS} &  \textbf{InfoGain} & \textbf{All attr.} & \textbf{CFS} & \textbf{InfoGain} &\\ \midrule
    
    \multirow{2}{*}{\textbf{NB}}
                 & \atributosDS  & 0,832 &  0,764 & 0,804 & 0,830 & 0,760 & 0,800 &\\ 
                 & \indicadoresDS & 0,792 &  \textbf{0,863} & 0,852 & 0,790 & 0,860 & 0,850 &\\ \addlinespace

    \multirow{2}{*}{\textbf{MLP}}
                & \atributosDS    & 0,781 &  0,790 & \textbf{0,861} & 0,780 & 0,790 & 0,860 &\\
     			& \indicadoresDS & 0,852 &  0,850 & 0,850 &0,850 & 0,850 & 0,850 &\\ \addlinespace

    \multirow{2}{*}{\textbf{SVM}}
                  & \atributosDS    & 0,303 &  0,300 & 0,635 &0,550 & 0,540 & 0,580 &\\ 
     			  & \indicadoresDS  & \textbf{0,892} &  0,852 & 0,852 & 0,890 & 0,850 & 0,850 &\\ \addlinespace

    \multirow{2}{*}{\textbf{AdaBoost}}
                 & \atributosDS    & 0,732 &  0,770 & 0,740 &0,730 & 0,770 & 0,840 &\\
       			 & \indicadoresDS & \textbf{0,823} &  0,820 & 0,820 & 0,820 & 0,820 & 0,820 &\\ \addlinespace

    \multirow{2}{*}{\textbf{JRip}}
                 & \atributosDS    & 0,648 &  0,655 & 0,710 &0,650 & 0,650 & 0,710 &\\ 
     			 & \indicadoresDS & 0,740 &  \textbf{0,802} & 0,784 & 0,740 & 0,800 & 0,780 &\\ \addlinespace
     			
    \multirow{2}{*}{\textbf{J48}}
                 & \atributosDS    & 0,689 &  0,700 & 0,704 &0,690 & 0,700 & 0,700 &\\
     			 & \indicadoresDS & 0,695 &  \textbf{0,800} & 0,780 & 0,690 & 0,800 & 0,780 &\\ \addlinespace

    \multirow{2}{*}{\textbf{RF}}
                & \atributosDS    & 0,840 &  0,790 & 0,795 &0,840 & 0,790 & 0,790 &\\ 
     			& \indicadoresDS & 0,891 &  \underline{\textbf{0,940}} & 0,920 &0,890 & 0,940 & 0,920 &\\ \addlinespace
    \bottomrule
\end{tabular}
\label{tab:multicol}
\end{table}

The algorithm Na\"ive Bayes (NB) presented its best result when CFS was employed over the \indicadoresDS\ data set, with a precision score of $0,863$. With InfoGain over the same data set, the best result was $0,852$ of precision. The worst result obtained with CFS was running it over the data set \atributosDS\, scoring a precision score of $0,764$. It has a probabilistic approach and assumes that the features are independents with the target value. This is not all true since that some indicators have influence in another one. For example, the business management influences the indicators of the sub-index Economic Balance. 

For Multilayer Percepton (MLP) its best result was obtained using InfoGain, presenting a precision score of $0,861$ for the \atributosDS\ data set. For \indicadoresDS~the results averaged $0,850$ of precision.

The Support Vector Machine (SVM) had its best results being executed on the \indicadoresDS\ data set and without the attributes selection, scoring $0,892$ of precision. With the data set \atributosDS\ the results were unsatisfactory, with $0,303$ of precision without attributes selection and $0,605$ with CFS. It performed better than the MLB and NB even without the need of the extra step of CFS or InfoGain. 

AdaBoost's has the best result in \indicadoresDS\ data set had $0,823$ of precision. With the data set \atributosDS\ the results didn't present many variations, and the best score was $0,770$ using CFS. This classifier uses a set of weak classifiers that combined creates the final model. Each weak classifier is a stamp that is tree with only two leaves called as stump. These stumps are weighted and combined reducing the error at each iteration. The AdaBoost has the advantage to be very fast with good result even with those simple classifiers. It can also provides the relevance of each feature, indicator in our case, in the final classification.

JRip presented its best result for the \indicadoresDS\ data set, with $0,802$ of precision for CFS. Using \atributosDS\, which is another data set, the best precision obtained was $0,710$, using InfoGain. One of the main advantages of this algorithm is that it produces readable rules as C4.5 rules and also is well fitted on continuous datasets~\cite{cohen1995fast} like our case. Comparing it with J48 can been that the result were closer each other. 
Using Random Forest (RF) the best result was obtained using the \indicadoresDS~data set and CFS, with $0.940$ of precision. With \atributosDS\ the best precision was obtained without the attributes selection, with $0,840$ of precision. The RF uses a combination of classification trees to produces the classification of the model. This technique generalizes very well delivering consistent good results in new data sample. The only disadvantage of it in our case is hard to check the rules that produces a result.

Thus, the Random Forest (RF) algorithm presents the best performance (94.0\%) of all algorithms, using seven attributes selected by CFS feature selection method: \textit{4. Degree of indebtedness}, \textit{8. Business management}, \textit{9. Information management}, \textit{12. Soil fertility}, \textit {16. Conservation practices}, \textit{19. APPs} (Adequacy of Permanent Preservation Areas) and \textit{20. Legal Reserve (Legal Reserve)}. This result is important because with only 7 of the 21 indicators we can infer with 94\% of precision the sustainability level of a rural property.

\section{Conclusion}

\label{cha:conclusion}

Public awareness of the negative impacts of human activity on our environment is at an all times high. Technological efforts to increase the sustainability of productive Agroecosystems are being studied, developed and applied in many different places. In this work, we adopt a Brazilian methodology called Indicators of Sustainability in Agroecosystems (\textit{Indicadores de Sustentabilidade em Agroecossistemas} -- ISA), implement an information system based on it and apply Data Science techniques over the gathered data - from 100 real rural properties - to compute which are the most relevant ISA Indicators for the final ISA Sustainability Index Score.

Initially, the ISA methodology for the calculation of sustainability in agroecosystems was presented. Based on this reference methodology, \isadigital~was developed. This new tool makes it possible for the methodology to be applied in a greater scale, allowing for quicker evaluation of participating properties, regarding their agroecosystems' sustainabilities.

One of the first contributions of the \isadigital~system was the structuring and organization of the information that is collected. This is one of the most important factors of this project. Without this organization and structuring, the development of this work would be much more complex, and further work using the data more difficult.

In order to expedite the collection of data on rural properties, a tool was developed for the use of agricultural technicians. The tool used by the technicians facilitates the work of these professionals, reducing the time needed to collect information, making it more difficult for human errors to happen and taking advantage of the platforms it runs on top to increase data security and integrity (Java and PostgreSQL, the traditional application of the ISA methodology involved Excel sheets). Security and integrity are possible by storing the questionnaires in a structured database, having the access controlled by the profile and access level of each user.

A web tool was also developed for a managerial audience, with a different visualization level and capabilities than the interfaces developed for agricultural technicians. It is possible to create projects for sustainability assessment using the system, and so technicians can submit questionnaires and their results under some project's umbrella, composing the project's data set. In these projects, technicians (responsible for collecting data) and project managers (who coordinate the work of the technicians allocated to the project) can be allocated.

Using the web tool, technicians can view reports of the rural properties they registered and submitted data from. With the aid of such a report, it is possible to elaborate an Adequation Plan, containing actions that must be implemented on a property for correction of failures and improvement of the processes, aiming to increase an agroecosystem's sustainability.

Project managers have access to reports in the same way as agricultural technicians, but also have access to a macro view of the projects they are responsible for. Managers have access to a managerial report and can visualize summarizations of the situation for a set of properties. Thus it is possible to see, for example, in a single report all the indicators of a city, sub-basin, region or year. These filters can be combined with each other to improve understanding of results and facilitate the management of actions that can be developed.

With the use of data mining techniques, it was possible to identify that using only 7 out of the 21 indicators - originally required by the ISA Model - it is possible to identify with 94\% precision the level of sustainability of a rural property. For this, the Random Forest classification technique was used. The indicators used are: \textit{4. Degree of indebtedness}, \textit{8. Business management}, \textit{9. Information management}, \textit{Soil fertility}, \textit {Conservation practices}, \textit{19. APPs} (Adequacy of Permanent Preservation Areas) and \textit{20. Legal Reserve (Legal Reserve)}. These indicators were selected using the CFS feature selection technique.

Some of the future work involves integration with new databases. Data can be collected from IBGE, such as data sets that can be useful to identify new variables with relevant correlations regarding sustainability. The National Water Agency (ANA) has updated maps of the basins and sub-basins of the country. This information can be cross-checked and be used to try to predict possible water scarcity for rural properties in some geographical region, for example. The integration of databases with railways, waterways and highways can help identify new production outflows routes or facilitate the creation of producers networks. Another future work possible, because demand was identified, is the development of a mobile application for georeferenced photographic registration of problems or solutions found in a property.

\begin{acks}
This work was partially funded by FAEMG, FAPEMIG, CAPES and CNPq.
\end{acks}

\bibliographystyle{ACM-Reference-Format}
\bibliography{bibfile}

\appendix

\section{ISA questionnaire fields}
\label{ch:appendix}

This appendix lists the names and enumeration of the fields in the questionnaire defined by the ISA Methodology. An instance of said questionnaire is fulfilled for each participating property. The questionnaire is composed by three sections: Questionnaire, Geoprocessing and Indicators.

\subsection{Part one: Questionnaire}

1. DATE OF INTERVIEW

2. IDENTIFYING THE INTERVIEWER:
 2.1 Name
; 2.2 CPF.

3. LOCATION AND IDENTIFICATION OF THE RURAL PROPERTY:
 3.1 Geographical coordinates of the rural property - GPS (headquarter or an identifiable reference point in the sketch): Degree, Latitude, Longitude, Altitude
; 3.2 Name of the municipality
; 3.3 Name of nearest watercourse
; 3.4 Code of the rural property.

4. IDENTIFYING THE INTERVIEWER:
 4.1 Name of the interviewee
; 4.2 CPF.

5. LAND OWNERSHIP TYPE

6. PRODUCER PROFILE:
 Owner's age.

7. DESCRIPTION OF RURAL PROPERTY:
 7.1 Name of the rural property
; 7.2 Description of the land(s) that make up the rural property (launch only contiguous areas = CAR)
; 7.3 Framing of the rural property, Size of the fiscal module in the municipality (ha)
; 7.4 Description of other areas not contiguous to the rural property and / or rental areas that integrate the income of the producer.

8. DESCRIPTION OF EMPLOYEES / PARTNERSHIPS:
 Workers / partnerships (Number): Permanent, Temporary, Sharecropper (including family members with direct link with production), Service exchange.

9. RESIDENCES IN THE RURAL PROPERTY:
 9.1 Number of family residences of the producer
; 9.2 Number of employee residences and / or sharecroppers.

10. SOIL USE AND OCCUPANCY ON THE RURAL PROPERTY:
 10.1 Specifications of production areas in the rural property (description, area, irrigation systems): Permanent crops, Temporary crops, Pastures, Forestry
; 10.2 Current and historical land use and occupation: Description, Current use Area (ha), proportion to total area (\%), Historic Area (ha) for Permanent crops, Temporary crops, Grassland, Forestry, Fallow land, Non-agricultural area, Water mirror, Native vegetation, Others.

11. ESTIMATED GROSS INCOME INSIDE AND OUT OF RURAL ENTERPRISE:
 11.1 Estimated annual gross income (from the rural enterprise), Activities / Products (values)
; 11.2 Estimated annual gross income (outside the rural enterprise), amounts: Pension, retirement, financial aid (grants and others) / Other activities / services.

12. ASSESSMENT OF RURAL PROPERTY (Quantities, values, historical values):
 12.1 Facilities and other betterments
; 12.2 Machinery and Equipment
; 12.3 Animals (bulls) Bulls, Oxen, Cows, Tourinhos and / or clubs, Heifers, Calves, Heifers, Equines, Muares, Pigs, Goats / Sheep, Poultry
; 12.4 Irrigation (table with name and area).

13. ESTIMATE OF THE VALUE OF THE RURAL PROPERTY:
 13.1 Land reference value in the region (R\$ / ha)
; 13.2 History - Land value in the region
; 13.3 Total estimated value of the rural property.

14. WATER RESOURCES:
 14.1 Number of springs and perennial water eyes in the rural property
; 14.2 Number of natural lakes and ponds in the rural property
; 14.3 Number of dams in rural property
; 14.4 Number of water courses in rural property
; 14.5 Source of water used in the establishment
; 14.6 Problems of water availability (for human consumption and activities).

15. RESIDUES AND EFFLUENTS GENERATED IN THE RURAL PROPERTY:
 15.1 Destination of sewage generated in residences
; 15.2 Destination of garbage (domestic and activities).

16. ENVIRONMENTAL REGULARIZATION OF THE RURAL PROPERTY:
 16.1 Has regularization of water use (grant or insignificant use)
; 16.2 Has environmental license or non-passable certificate or AAF
; 16.3 It has regularization of the Legal Reserve and Permanent Preservation Areas.

17. CRITICAL POINTS OF THE ENTERPRISE

18. EVALUATION OF WATER QUALITY IN THE RURAL PROPERTY:
 18.1 Type of occupation of the banks of the water body (main activity)
; 18.2 Antropic changes
; 18.3 Shading from the vegetal cover in the bed (from the margins)
; 18.4 Near erosion and / or on the banks of the body of water and silting in its bed
; 18.5 Water transparency
; 18.6 Odor of water
; 18.7 Water Oilyness
; 18.8 Sediment odor (background)
; 18.9 Sediment oiliness (bottom)
; 18.10 Type of background.

\subsection{Part Two: Geoprocessing}

2. USE AND OCCUPATION OF SOIL IN THE RURAL PROPERTY:
 2.1 - Land use and occupation according to CAR (ha) sketch: Cropland / Grassland / Forestry / non-agricultural area, Fallowing Area, Water Mirror (reservoirs) and watercourses, Remnant of Native Vegetation.
; 2.2 - Administrative easement.

3. AREAS OF PERMANENT PRESERVATION (APPs) IN THE RURAL PROPERTY:
 3.1 - Wet and dry APPs Area (ha) (\%), wet APPs, dry APPs, TOTAL
; 3.2 - Land use in PPAs: Area (ha), proportion to total area (\%): Native vegetation, Area to be reclaimed, Area of use consolidated, Area to be reclaimed for real estate up to 4MF.

4. REMANESCENTS OF NATIVE VEGETATION IN THE RURAL PROPERTY:
 4.1 - Native vegetation outside APPs, Native vegetation area (ha)
; 4.2 - Area with native vegetation exceeding the area required for RL in the rural property, Area of native vegetation (ha).

\subsection{Part three: Indicators}

1. PRODUCTIVITY INDICES AND CLEARING SALES, Main activities of the establishment:
 1.1 Description of activities
; 1.2 Units of measurement
; 1.3 Current average productivity
; 1.4 Average selling price (R\$ / un.)
; 1.5 Average productivity in the region
; 1.6 Average price of the region (R\$ / un.).
       
2. INCOME DIVERSITY: Income share (\%) Weighting factor:
 2.1. Agricultural, livestock and forestry activities
; 2.2 Other activities in the establishment: tourism, handicrafts, agribusiness
; 2.3 Other off-premises activities
; 2.4 Retirement; Pension; Financial help; Other sources of income
; 2.5 Verification - occurrence of concentration of agricultural income in a single activity (> 80\% of total income verified inside and outside the rural enterprise).

3. RURAL PROPERTY ASSETS DEVELOPMENT: Evolution in the period (
 3.1 Land value in the region
; 3.2 Improvements
; 3.3 Equipment
; 3.4 Livestock
; 3.5 Extension of the area of irrigated agriculture and / or agriculture
; 3.6 Balance Sheet
; 3.7 Balance sheet not counting on the valuation of land.
       
4. DEGREE OF INDEBTEDNESS:
 4.1 Debt value in relation to equity (\%).

5. BASIC SERVICES AVAILABLE FOR RURAL PROPERTY / FOOD SAFETY:
 5.1 Basic services available in the residences: Availability of water (quantity and quality), Access to electricity, Regular access to production and receipt of inputs, Access to health service, Regular access to school transport, Field security (patrol for rural policing), Telephone (cellular, rural or fixed cellular), Internet, Public garbage collection
; 5.2 Food Security (directed to establishments classified as Family Agriculture). Check the surroundings of the residences (own family consumption): Vegetables, grains and tubers, Orchard, Breeding of animals.

6. SCHOOLING / COURSES ADDRESSED TO THE MAIN ACTIVITIES IN THE RURAL PROPERTY:
 6.1 Number of people in the establishment
; 6.2 Less than 5 years of study
; 6.3 5 to 9 years of study
; 6.4 Over 9 years of study
; 6.5 Upper course
; 6.6 Training short season
; 6.7 Long-term training
; 6.8 Attend school network.
       
7. QUALITY OF OCCUPATION AND EMPLOYMENT GENERATED:
 7.1 Total of people in the rural property
; 7.2 Employee registration (work permit)
; 7.3 Overtime payment (or hour bank)
; 7.4 Over 1 minimum wage
; 7.5 Feeding aid
; 7.6 Housing assistance
; 7.7 Education and transport aid
; 7.8 Profit Sharing
; 7.9 Accident insurance
; 7.10 Access to leisure
; 7.11 Space for growing food.
       
8. BUSINESS MANAGEMENT:
 8.1 Cash flow (income / expense)
; 8.2 Cost of production of activities
; 8.3 Access to technical assistance (private or public)
; 8.4 Participation - associative forms - active (1) or passive (0.5)
; 8.5 Environmental regulation (water use, RL, APP and licensing)
; 8.6 Use credit for investment
; 8.7 Use credit for costing
; 8.8 Uses credit for commercialization.
       
9. INFORMATION MANAGEMENT:
 9.1 Seeks information for commercialization of the production / seeks diversification of buyers
; 9.2 Generates certified products and / or institutional market
; 9.3 Adoption of Innovative Techniques
; 9.4 Capacity for innovation or leadership in the community.
       
10. MANAGEMENT OF RESIDUES AND EFFLUENTS GENERATED IN THE RURAL PROPERTY:
 10.1 Proper collection and disposal of waste (household and business waste, recyclable and non-recyclable)
; 10.2 Appropriate disposal of domestic sewage
; 10.3 Composting and / or reuse of organic solid waste
; 10.4 Adequate disposal and treatment of liquid effluents (generated by creations or beneficiation units) (Use of treated effluents in production systems).
; 10.5 Treatment of gaseous effluents (generated in boilers, biodigesters, charcoal).
; 10.6 The rural property has some critical point regarding the management of residues and effluents.
      
11. WORK SAFETY / MANAGEMENT OF THE USE OF AGROCHEMICALS AND VETERINARY PRODUCTS:
 11.1 How many people handle agrochemicals and / or veterinary products
; 11.2 How Many People Use PPE Properly
; 11.3 Proper storage of packaging
; 11.4 Return of pesticide containers and correct destination of packaging of veterinary products.

12 SOIL PRODUCTIVE CAPACITY - FERTILITY:
 12.1 SOIL TEXTURE
; 12.2 ORGANIC MATTER
; 12.3 PHOSPHORUS AVAILABLE
; 12.4 EXCHANGEABLE CALCIUM
; 12.5 CHANGEABLE MAGNESIUM
; 12.6 CHANGEABLE POTASSIUM
; 12.7 ACTIVE ACIDITY (pH)
; 12.8 CHOCOLATE ALUMINUM
; 12.9 EFFECTIVE CTC
; 12.10 SATURATION BY BASES.

13 WATER QUALITY ASSESSMENT:
 13.1 - SURFACE WATER: Water pH, Thermotolerant Coliforms, Turbidity, Total Nitrate
; 13.2 - UNDERGROUND WATER / HUMAN CONSUMPTION: Water pH, Thermotolerant Coliforms, Total Nitrate.

14 RISK OF WATER CONTAMINATION BY AGROCHEMISTS IN THE RURAL PROPERTY:
 14.1 - Trade name of the product
; 14.2 - Description of crop or field
; 14.3 - Applied area (ha)
; 14.4 - Volume applied L / ha or kg / ha
; 14.5 - Active principle of the product
; 14.6 - Toxicity to fish - 96h LC50 (mg L-1)
; 14.7 - Toxicity - DAH (mg kg -1 day -1)
; 14.8 - Koc
; 14.9 - t1 / 2 (DT50)
; Clay content in the soil> 60\% (1), 30\% - 60\% (2), <30\% (3)
; 14.11 - Distance to water course (edge of field)> 1,000m (1), 300 - 1,000m (2), <300m (3)
; 14.12 - Soil management type: Protected soil (1), Soil without tillage (2) ;, Soil with tillage (3)
; 14.13 - Contamination risk (1, 2 or 3)
; 14.14 - Identification of the field
; 14.15 - Area (ha)
; 14.16 - Maximum risk (1), (2) or (3)

15 EVALUATION OF AREAS WITH SOIL IN THE PROCESS OF DEGRADATION IN THE RURAL PROPERTY:
 15.1 - Intensity of degradation stage
; 15.2 - Trend of process behavior.

16 DEGREE OF ADOPTION OF CONSERVATION PRACTICES IN THE RURAL PROPERTY:
; 16.1 - Degree of adoption of practices for soil conservation
 16.2 - Degree of adoption of strategies for the conservation and reservation of water in the rural property.

17 STATE OF CONSERVATION OF ROADS THAT CUT AND MARK THE RURAL PROPERTY:
17.1 - Presence of conservation and drainage system on roads: Transversal declivity of roads, Presence of bumps / furrows for runoff diversion, Presence of infiltration boxes
; 17.2 - Preservation of roads Presence of holes in roads: Presence of erosion grooves on roads
; 17.3 - The rural property has some critical point on the roads.

18 NATIVE VEGETATION - VEGETATION TYPES AND STATE OF CONSERVATION IN THE RURAL PROPERTY:
 18.1 - Sucessional stages of native vegetation: Advanced stage, Medium stage, Initial stage
; 18.2 - Proportion of the area of the fragments with protected native vegetation * Advanced stage, Medium stage, Initial stage
; 18.3 - Number of fragments with native vegetation (in the rural property)
; 18.4 - This fragment (s) has connection with the native vegetation of neighboring properties.

19 ADEQUACY OF PERMANENT PRESERVATION AREAS (APPs) OF THE RURAL PROPERTY:
 19.1 - Soil use and occupation in PPAs Native vegetation: Anthropogenic area with consolidated use that can be exploited, Vegetation suppression area with obligation to recompose vegetation. native
; 19.2 - Proportion of PPPs effectively protected and with good conservation status (

20 ADEQUACY OF THE LEGAL RESERVE (RL) OF THE RURAL PROPERTY:
20.1 - Adequacy of RL: Native vegetation surplus to RL, RL outside rural property, Areas from / to RL not subject to recomposition, Areas from / to RL that require recomposition.

21 DIVERSIFICATION OF THE AGROSSILVIPASTORIL LANDSCAPE IN THE RURAL PROPERTY:
 21.1 - Degree of adoption of practices that aid in agrobiodiversity
; 21.2 - Shanon Index (productive areas and native vegetation)
; 21.3 - There is a high proportion (more than 70\% of the perimeter) of monocultures around the rural property.
\clearpage

\section{\isadigital: Human interaction and Data delivery Process}
\label{ch:appendix2}

\begin{figure}[h]
  \centering
  \includegraphics[width=0.71\linewidth]{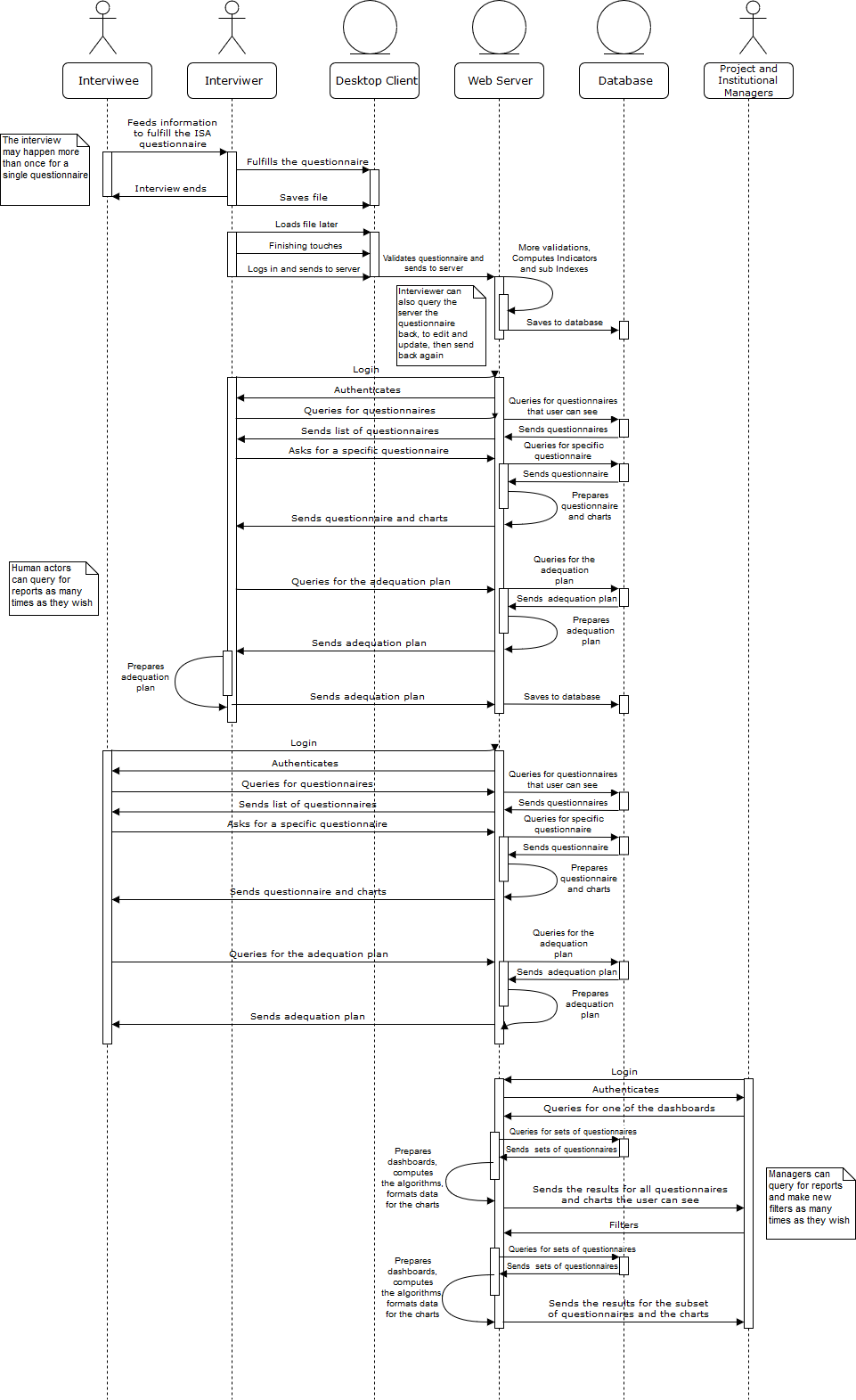}
  \caption{The UML diagram for the most relevant human interaction and questionnaires delivery process of the \isadigital~ system (\url{https://agro.sybers.dcc.ufmg.br/}).}
  \label{fig:AProcess}
\end{figure}

\end{document}